\documentclass[journal]{IEEEtran}
 
\usepackage{epsfig}

\usepackage{amsthm}
\usepackage[cmex10]{amsmath}
\usepackage{graphicx}
\usepackage{times}
\usepackage{latexsym}
\usepackage{graphicx}
\usepackage{amssymb}
\usepackage{amsfonts}
\usepackage{psfrag}

\newtheorem{thm}{Theorem}[section]
\newtheorem{cor}[thm]{Corollary}
\newtheorem{lem}[thm]{Lemma}

\newtheorem{definition}{Definition}
\theoremstyle{remark} \newtheorem{stdseq}{Standard LDPC Degree Distribution}
\theoremstyle{remark} \newtheorem{tetseq}{Two Edge Type Degree Distribution}
\theoremstyle{remark} \newtheorem{example}{Example}

\newcommand{\brc}[1]{\left( #1 \right)}
\newcommand{\coef}[2]{\textnormal{coef}\left\{#1,#2\right\}}
\newcommand{\ul}[1]{\underline{#1}}

\newcommand{\openone}{\leavevmode\hbox{\small1\normalsize\kern-.33em1}}
\begin{document}

\title{Performance Analysis and Design of Two Edge Type LDPC Codes for the BEC Wiretap Channel}



\author{Vishwambhar~Rathi, Mattias~Andersson,~\IEEEmembership{Student~Member,~IEEE,} Ragnar~Thobaben,~\IEEEmembership{Member,~IEEE,} Joerg~Kliewer,~\IEEEmembership{Senior~Member,~IEEE,} Mikael~Skoglund,~\IEEEmembership{Senior~Member,~IEEE}%
\thanks{V. Rathi is with Nvidia Corporations. This work was done when he was with the School of Electrical Engineering and 
the ACCESS Linnaeus Centre, KTH Royal Institute of Technology, Stockholm, Sweden (email:vrathi@gmail.com).}
\thanks{M. Andersson, R. Thobaben, and M. Skoglund are with the School of Electrical Engineering and the ACCESS Linnaeus Centre, KTH Royal Institute of Technology, Stockholm, Sweden (e-mail: {amattias, ragnar.thobaben, skoglund}@ee.kth.se).}%
\thanks{J. Kliewer is with the Klipsch School of Electrical and Computer Engi- neering, New Mexico State University, Las Cruces, NM 88003, USA (e-mail: jkliewer@nmsu.edu).}%
\thanks{Part of this work has appeared as
conference papers in \cite{RATKS09, ARTKS10asl}.}
\thanks{This work has been supported in part by U.S. National Science Foundation grants CCF-0830666 and CCF-1017632.}}

\maketitle
\begin{abstract}
We consider transmission over a wiretap channel where both the main channel and
the wiretapper's channel are Binary Erasure Channels (BEC).  We propose a code
construction method using two edge type Low-Density Parity-Check (LDPC) codes
based on the coset encoding scheme. 
Using a standard LDPC ensemble with a given threshold over the BEC, we give a
construction for a two edge type LDPC ensemble with the same threshold. If the
given standard LDPC ensemble has degree two variable nodes, our construction
gives rise to degree one variable nodes in the code used over the main channel.
This results in zero threshold over the main channel. In order to circumvent
this problem, we numerically optimize the degree distribution of the two edge type
LDPC ensemble.  We find that the resulting ensembles are able to perform close to the boundary of the rate-equivocation
region of the wiretap channel. 

There are two performance criteria for a coding scheme used over a wiretap
channel: reliability and secrecy. The reliability measure corresponds to the
probability of decoding error for the intended receiver. This can be easily
measured using density evolution recursion.  However, it is more challenging to
characterize secrecy, corresponding to the equivocation of the message for the
wiretapper.  M\'{e}asson, Montanari, and Urbanke have shown how the equivocation
can be measured for a broad range of standard LDPC ensembles for transmission
over the BEC under the point-to-point setup. By  generalizing the method of
M\'{e}asson, Montanari, and Urbanke to two edge type LDPC ensembles, we show how
the equivocation for the wiretapper can be computed.  We find that relatively
simple constructions give very good secrecy performance and are close to the
secrecy capacity. However finding explicit sequences of two edge type LDPC ensembles which
achieve secrecy capacity is a more difficult problem. We pose it as an
interesting open problem. 

\end{abstract}
 \IEEEpeerreviewmaketitle
\section{Introduction}\label{sec:introduction}
\begin{figure}[htbp]
  \centering
  \includegraphics[width=\columnwidth]{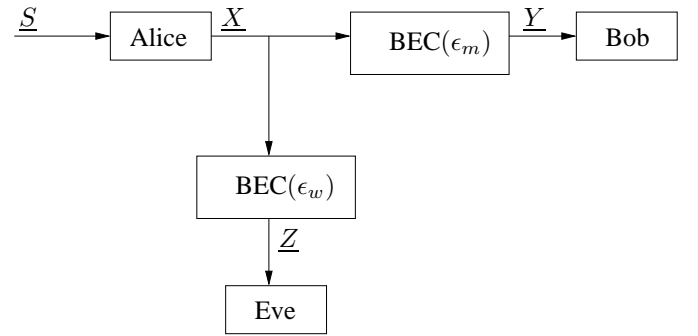}
  \caption{A BEC wiretap channel.}
  \label{fig:channel}
\end{figure}

Wyner introduced the notion of a wiretap channel in \cite{Wyn75} which is
depicted in Figure \ref{fig:channel}.  In general, the channel from Alice to
Bob and the channel from Alice to Eve can be any discrete memoryless channels.
In this paper we will restrict ourselves to the setting where both channels are
Binary Erasure Channels (BEC). We denote a BEC with erasure probability
$\epsilon$ by BEC($\epsilon$). In a wiretap channel, Alice communicates a
message $\ul{S}$, which is chosen uniformly at random from the message set $\mathcal{S}$, to Bob through the
main channel which is a BEC($\epsilon_m$). Alice performs this task by encoding
$\ul{S}$ as an $n$ bit vector $\ul{X}$ and transmitting $\ul{X}$ across
BEC($\epsilon_m$). Bob receives a noisy version of $\ul{X}$ which is denoted by
$\ul{Y}$. Eve observes $\ul{X}$ via the wiretapper's channel BEC($\epsilon_w$)
and receives a noisy version of $\ul{X}$ denoted by $\ul{Z}$. We denote such a
wiretap channel by BEC-WT($\epsilon_m, \epsilon_w$). 

\psfrag{Alice}{Alice}
\psfrag{Bob}{Bob}
\psfrag{Eve}{Eve}
\psfrag{W}{$\ul{S}$}
\psfrag{X}{$\ul{X}$}
\psfrag{Y}{$\ul{Y}$}
\psfrag{Z}{$\ul{Z}$}
\psfrag{BECm}{BEC$(\epsilon_m)$}
\psfrag{BECw}{BEC$(\epsilon_w)$}

The encoding of a message $\ul{S}$ by Alice should be such that Bob is able to decode
$\ul{S}$ reliably and $\ul{Z}$ provides as little information as possible to Eve about $\ul{S}$.


%
%

A detailed information theoretic overview of general wiretap channels can be
found in \cite{LiPS09}. In \cite{Thangaraj}, \cite{Poor2} the authors have
given code design criteria using sparse graph codes. Their approach is based
on a coset coding scheme using nested codes \cite{ZSE02}. 
In \cite{SSTBM10} the authors have suggested a coding scheme for the BEC-WT
that guarantees strong secrecy for a noiseless main channel and some range of
$\epsilon_w$ using duals of sparse graph codes. In \cite{ChV10} it was shown
that random linear codes can achieve the secrecy capacity over the binary
symmetric wiretap channel and an upper bound on the information leakage was
derived.  Recently it has been shown that using Arikan's polar codes
\cite{Ari09}, it is possible to achieve the whole rate-equivocation region
\cite{ARTKS10}, \cite{MaV10}, \cite{OzG10}, \cite{HoS10}.

We propose a code construction method using two edge type LDPC codes based on
the coset encoding scheme. The {\it threshold} of a code (or an ensemble) for
transmission over the BEC is the largest erasure probability for which reliable
communication is possible. Using a standard LDPC ensemble with a given
threshold over the BEC, we give a construction for a two edge type LDPC ensemble
with the same threshold. Thus if the standard LDPC ensemble is capacity
achieving over the wiretapper's channel, our construction of the two edge type
LDPC ensemble guarantees perfect secrecy.  Hence it achieves secrecy capacity
if $\epsilon_m=0$ i.e. the main channel is noiseless.  

However, our construction cannot guarantee reliability over the main channel
if $\epsilon_m > 0$ and the given standard LDPC ensemble has degree two
variable nodes. This is because our approach gives rise to degree one
variable nodes in the code used over the main channel.  This results in zero
threshold over the main channel. In order to circumvent this problem, we
numerically optimize the degree distribution of the two edge type LDPC
ensemble.  We find that the resulting codes approach the rate-equivocation
region of the wiretap channel.  For example, for the BEC-WT($0.5, 0.6$) we find
ensembles that achieve the points $(R_{ab},R_e) = (0.0999064,0.0989137)$ and
$(R_{ab},R_e) = (0.498836,0.0989137)$ which are very close to the best achievable points
$B = (0.1,0.1)$ and $C = (0.5,0.1)$ as depicted in Figure \ref{fig:rate_equ}.  
The definitions of $R_{ab}$, $R_e$, and a description of Figure \ref{fig:rate_equ} 
 are given in Section \ref{sec:codeconst}.

Note that reliability, which corresponds to the probability of decoding error
for the intended receiver, can be easily measured using density evolution
recursion. However secrecy, which is given by the equivocation of the message
conditioned on the wiretapper's observation, can not be easily calculated.
M\'{e}asson, Montanari, and Urbanke have derived a method to measure equivocation
for a broad range of standard LDPC ensembles for point-to-point transmission
over the BEC \cite{MMU08}. From now onwards we call it the MMU method\footnote{We call it the MMU method in acknowledgment of 
M\'{e}asson, Montanari, and Urbanke, the authors of \cite{MMU08}.}. The MMU method was extended to
non-binary LDPC codes for transmission over the BEC in \cite{Vra08, RaA10}.  By
generalizing the MMU method for two edge type LDPC ensembles, we show how the
equivocation for the wiretapper can be computed.  We find that relatively
simple constructions give very good secrecy performance and are close to the
secrecy capacity.

Our paper is organized in the following way. In Section \ref{sec:codeconst}, we
give various definitions, describe the coset encoding method and two edge type
LDPC ensembles, and give the density evolution recursion for two edge type LDPC
ensembles.   Section \ref{sec:desandopt} contains the code design and
optimization for the BEC wiretap channel BEC-WT($\epsilon_m, \epsilon_w$). In
Section \ref{sec:compequ}, we show that the task of computing the equivocation
is equivalent to generalizing the MMU method for two edge type LDPC ensemble for 
point-to-point transmission over the BEC. We 
generalize the MMU method for two edge type LDPC ensemble in Section
\ref{sec:mmu_tet}.  In Section \ref{sec:examples} we present various examples
to elucidate the computation of equivocation and show that our optimized degree
distributions also approach the information theoretic equivocation limit.
Finally, we conclude in \ref{sec:conc} with some discussion and open problems.

\section{Code Construction}\label{sec:codeconst}
\begin{figure}[htp]
\centering
\setlength{\unitlength}{1bp}%
\begin{picture}(113,90)
\put(0,0){\includegraphics[scale=0.9]{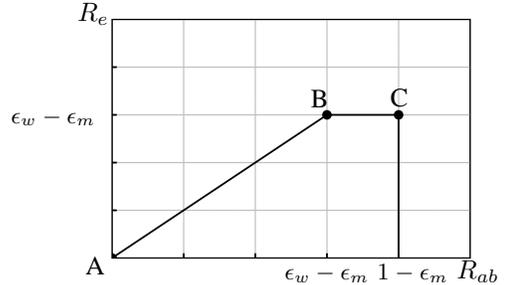}}
{
\put(-13, 88){\makebox(0,0)[lb]{$R_e$}}
\put(-10, -6){\makebox(0,0)[lb]{A}}
\put(130,-9){\makebox(0,0)[lb]{$R_{ab}$}}
\put(100,-9){\makebox(0,0)[lb]{\small $1-\epsilon_m$}}
\put(65,-9){\makebox(0,0)[lb]{\small $\epsilon_w-\epsilon_m$}}
\put(-38, 50){\makebox(0,0)[lb]{\small $\epsilon_w-\epsilon_m$}}
\put(75, 57){\makebox(0,0)[lb]{B}}
\put(105, 57){\makebox(0,0)[lb]{C}}
}
\end{picture}
\caption{\label{fig:rate_equ} Achievable rate equivocation region for BEC-WT($\epsilon_m, \epsilon_w$).} 
\end{figure}

We first define a code for the wiretap channel. 
\begin{definition}[Code for Wiretap Channel]
A code of rate $R_{ab}$ with block length $n$ for the wiretap channel is given
by a message set $\mathcal S$ of cardinality $|\mathcal{S}| = 2^{n R_{ab}}$,
and a set of disjoint sub-codes $\{\mathcal C(\ul{s}) \subset \mathcal
X^n\}_{\ul{s} \in \mathcal S}$. 
messages.  To encode the message $\ul{s} \in \mathcal S$, Alice chooses one of
the codewords in $\mathcal C(\ul{s})$ uniformly at random and transmits it. Bob
uses a decoder $\phi: \mathcal Y^n \to \mathcal S$ to determine which message
was sent.
\end{definition}
We now define the achievability of rate of communication from Alice to Bob and equivocation 
of the message from Alice to Bob for Eve. 
\begin{definition}[Achievability of Rate-Equivocation]
A rate-equivocation pair $(R_{ab},R_e)$ is said to be achievable if $\forall \epsilon > 0$, 
there exists a sequence of codes of rate $R_{ab}$ of length $n$ and decoders $\phi_n$ such that the 
following reliability and secrecy criteria are satisfied. 
\begin{align}
\label{eq:rate}
\textnormal{Reliability:} \lim_{n \to \infty} \quad P(\phi_n(\ul{Y}) \neq \ul{S}) < \epsilon,
\end{align}
\begin{align}
\textnormal{Secrecy:} \liminf_{n \to \infty} \frac 1 n H(\ul{S}|\ul{Z}) > R_e - \epsilon \label{eq:security}.
\end{align}
\end{definition}
Note that we use the weak notion of secrecy as opposed to the strong notion \cite{LiPS09}.
With a slight abuse of terminology, when we say equivocation we mean the
normalized equivocation as defined in the LHS of (\ref{eq:security}). From the achievable 
rate-equivocation region for general wiretap channels given in   
\cite{Wyn75}, the set of achievable pairs $(R_{ab},R_e)$ for the BEC-WT($\epsilon_m,
\epsilon_w$) is given by 
\begin{equation}\label{eq:ach}
   R_e \leq R_{ab} \leq 1 - \epsilon_m, \quad 0 \leq R_e \leq \epsilon_w - \epsilon_m. 
\end{equation}
The rate region described by (\ref{eq:ach}) is depicted in Figure \ref{fig:rate_equ}.

The line segment AB in Figure \ref{fig:rate_equ} corresponds to to perfect secrecy.  
\begin{definition}[Perfect Secrecy and Secrecy Capacity \cite{Wyn75}]
The points in the achievable region where $R_{ab}=R_e$ correspond  to
{\it perfect secrecy} i.e. for these points $I(\ul{Z};\ul{S})/n \to 0$. The highest
achievable rate $R_{ab}$ at which we can achieve perfect secrecy is called the
{\it secrecy capacity}  and we denote it by $C_S$. 
\end{definition}
For the BEC-WT($\epsilon_w, \epsilon_m$), we have $C_S = \epsilon_w - \epsilon_m$.
We now describe the coset encoding and syndrome decoding method. Let
$H$ be an $n (1-R) \times n$ LDPC matrix. Let $\mathcal C$ be the code
whose parity-check matrix is $H$. Let $H_1$ and $H_2$ be the
sub-matrices of $H$ such that
 \begin{equation*}
  H = \begin{bmatrix}
    H_1 \\ H_2
  \end{bmatrix}, 
\end{equation*}
where $H_1$ is an $n (1-R_1) \times n$ matrix. Clearly, $R_1 > R$.
Let $\mathcal C_1$ be the code with parity-check matrix
$H_1$. $\mathcal C$ is the coarse code and $\mathcal C_1$ is the fine
code in the nested code $(\mathcal C_1,\mathcal C)$ \cite{ZSE02}. Also,
$\mathcal{C}_1$ is partitioned into $2^{n(R_1-R)}$ disjoint subsets
given by the cosets of $\mathcal C$. Alice uses
the {\it coset encoding method} to communicate her message to Bob which we now describe. \\
\begin{definition}[Coset Encoding Method] 
Assume that Alice wants to transmit a
message whose binary representation is given by an $n (R_1-R)$-bit
vector $\ul{S}$. To do this she performs coset encoding by transmitting $\ul{X}$, which is a
randomly chosen solution of
\[
   \begin{bmatrix}
    H_1 \\ H_2
  \end{bmatrix} \ul{X} = [0 \cdots 0 \ \ul{S}]^T. 
\]
\end{definition}
Bob uses the following {\it syndrome decoding} to retrieve the message from Alice. 
\begin{definition}[Syndrome Decoding]
 After observing $\ul{Y}$, Bob obtains an estimate $\hat{\ul{X}}$ for $\ul{X}$
using the parity check equations $H_1 \ul{X} = 0$. Then he computes an estimate 
$\ul{\hat{S}}$ for $\ul{S}$ as $\ul{\hat{S}} = H_2 \hat{\ul{X}}$, where $\ul{\hat{S}}$ is the syndrome of $\ul{\hat{X}}$ with respect to the matrix $H_2$.
\end{definition}

\begin{figure}[htbp]
  \centering
  \includegraphics[width=\columnwidth]{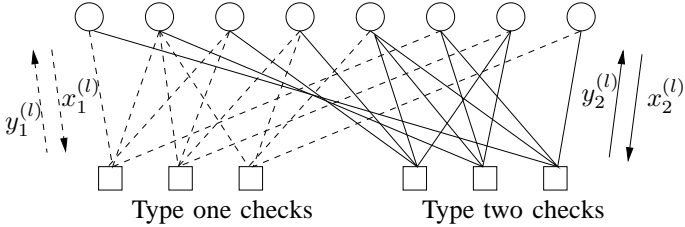}
  \caption{Two edge type LDPC code.}
  \label{fig:ldpc}
\end{figure}

A natural candidate for coset encoding is a two edge type LDPC code \cite{Mac99}. 
A two edge type matrix $H$ has form 
 \begin{equation}\label{eq:twoedgemat}
  H = \begin{bmatrix}
    H_1 \\ H_2
  \end{bmatrix}.  
\end{equation}
The two types of edges are the edges connected to check nodes in $H_1$ and those connected to check nodes in $H_2$. 
An example of a two edge type LDPC code is shown in Figure \ref{fig:ldpc}.
\psfrag{Type 1 checks}{Type one checks}
\psfrag{Type 2 checks}{Type two checks}
\psfrag{x1l}{$x_1^{(l)}$}
\psfrag{x2l}{$x_2^{(l)}$}
\psfrag{y1l}{$y_1^{(l)}$}
\psfrag{y2l}{$y_2^{(l)}$}

We now define the degree distribution of a two edge type LDPC
ensemble. Let $\lambda^{(j)}_{l_1 l_2}$ denote the fraction of type
$j$ ($j=1$ or $2$) edges connected to variable nodes with $l_1$
outgoing type one edges and $l_2$ outgoing type two edges. The fraction
$\lambda^{(j)}_{l_1 l_2}$ is calculated with respect to the total
number of type $j$ edges. Let $\Lambda_{l_1 l_2}$ be the fraction of
variable nodes with $l_1$ outgoing edges of type one and $l_2$
outgoing edges of type two. This gives the following relationships
between $\Lambda, \lambda^{(1)}$, and $\lambda^{(2)}$,
\begin{align}
  \label{eq:match1}
  \lambda^{(1)}_{l_1 l_2} &= \frac{l_1 \Lambda_{l_1 l_2}}{\sum_{i_1, i_2} i_1 \Lambda_{i_1 i_2}}, \\
  \label{eq:match2}
  \lambda^{(2)}_{l_1 l_2} &= \frac{l_2 \Lambda_{l_1 l_2}}{\sum_{i_1, i_2} i_2 \Lambda_{i_1 i_2}}, \\
  \label{eq:match3}
  \Lambda_{l_1 l_2} & = \frac{\frac{\lambda_{l_1 l_2}^{(1)}}{l_1}}{\sum_{i_1, i_2}\frac{\lambda_{i_1 i_2}^{(1)}}{i_1}} = \frac{\frac{\lambda_{l_1 l_2}^{(2)}}{l_2}}{\sum_{i_1, i_2}\frac{\lambda_{i_1 i_2}^{(2)}}{i_2}}.
\end{align}
Similarly, let $\rho^{(j)}_r$ and $\Gamma^{(j)}_r$ denote the degree
distribution of type $j$ edges on the check node side from the edge and node
perspective respectively. Note that only one type of edges is connected to a
particular check node. $\Gamma^{(j)}_r$ and $\rho^{(j)}_r$ are related as 
follows,
\begin{align}
\rho^{(j)}_r &= \frac{r \Gamma^{(j)}_r}{\sum_{i} i \Gamma^{(j)}_i}, \label{eq:Gammatorho}\\ 
\Gamma^{(j)}_r &= \frac{\frac{\rho^{(j)}_r}{r}}{\sum_{i} \frac{\rho^{(j)}_i}{i}}. \label{eq:rhotoGamma} 
\end{align}

An equivalent definition of the degree distribution is given by the following polynomials:
\begin{align}
  \Lambda(x,y) & =  \sum_{l_1,l_2} \Lambda_{l_1 l_2} x^{l_1} y^{l_2}, \label{eq:polLam}\\
  \lambda^{(1)}(x,y) &= \sum_{l_1, l_2} \lambda^{(1)}_{l_1 l_2} x^{l_1 - 1}y^{l_2}, \label{eq:pollam1}\\
  \lambda^{(2)}(x,y) &= \sum_{l_1, l_2} \lambda^{(1)}_{l_1 l_2} x^{l_1}y^{l_2 - 1}, \label{eq:pollam2}\\
  \Gamma^{(j)}(x) &= \sum_r \Gamma^{(j)}_r x^r, \quad j=1,2, \label{eq:polGamma}\\
  \rho^{(j)}(x) &= \sum_r \rho^{(j)}_r x^{r-1}, \quad j=1,2. \label{eq:polrho}
\end{align}
Like the standard LDPC ensemble of \cite{RSU01}, the two edge type LDPC
ensemble with block length $n$ and degree distribution $\left\{\lambda^{(1)},
\lambda^{(2)}, \rho^{(1)}, \rho^{(2)}\right\}$ ($\{\Lambda, \Gamma^{(1)},
\Gamma^{(2)}\}$ from node perspective) is the collection of all bipartite
graphs satisfying the degree distribution constraints, where we allow multiple
edges between two nodes. We will denote a {\it left regular} two edge type LDPC
ensemble for which $\Lambda(x, y) = x^{l_1} y^{l_2}$ by $\{l_1, l_2,
\Gamma^{(1)}, \Gamma^{(2)}\}$.

Consider the two edge type LDPC ensemble $\{\Lambda, \Gamma^{(1)},
\Gamma^{(2)}\}$.  If we consider the ensemble of the subgraph induced by
one particular type of edges then it is easy to see that the resulting
ensemble is the standard LDPC ensemble and we can easily calculate its
degree distribution. Let $\{\Lambda^{(j)}, \Gamma^{(j)}\}$ be the
degree distribution from node perspective ($\{\lambda^{(j)}, \rho^{(j)}\}$ 
from edge perspective) of the ensemble induced by type $j$ edges,
$j=1,2$. Then $\Lambda^{(j)}$, for $j=1, 2$, is given by
\begin{equation}\label{eq:Lam12}
\Lambda^{(1)}_{l_1} = \sum_{l_2} \Lambda_{l_1 l_2}, \quad \Lambda^{(2)}_{l_2} = \sum_{l_1} \Lambda_{l_1 l_2}. 
\end{equation}
The corresponding polynomials are defined as 
\begin{equation}\label{eq:polLam12}
\Lambda^{(1)}(x) = \sum_{i} \Lambda^{(1)}_{i} x^i, \quad \Lambda^{(2)}(x) = \sum_{i} \Lambda^{(2)}_{i} x^i. 
\end{equation}

To illustrate the relationship between various degree distributions, we consider 
a two edge type LDPC ensemble with degree distribution
\begin{align*}
\Lambda(x,y)  &= 0.2 x^3 y^4 + 0.4 x^3 y^5 + 0.4 x^6 y^6, \\ 
\Gamma^{(1)}(x) &= 0.6 x^7 + 0.4 x^8,\\
 \Gamma^{(2)}(x) &= x^{10}. 
\end{align*}
Using (\ref{eq:match1}), (\ref{eq:match2}), (\ref{eq:match3}),  
(\ref{eq:Gammatorho}), (\ref{eq:rhotoGamma}), and (\ref{eq:Lam12}), we obtain
\begin{align*}
\lambda^{(1)}(x,y) &= \frac{1}{7} x^2 y^4 + \frac{2}{7} x^2 y^5 + \frac{4}{7} x^5 y^6, \\
\lambda^{(2)}(x,y) &=  \frac{2}{13} x^3 y^3 +  \frac{5}{13} x^3 y^4 +  \frac{6}{13} x^6 y^5, \\
\rho^{(1)}(x) &= \frac{21}{37} x^6 + \frac{16}{37} x^7, \\
\rho^{(2)}(x) &= x^9, \\
\Lambda^{(1)}(x) &= 0.6 x^3 + 0.4 x^6, \\
\Lambda^{(2)}(x) &= 0.2 x^4 + 0.4 x^5 + 0.4 x^6.
\end{align*}

We now derive the density evolution equations for two edge type LDPC
ensembles, assuming that transmission takes place over the 
BEC($\epsilon$).  Let $x^{(l)}_j$ denote the probability that a
message from a variable node to a check node on an edge of type $j$ in
iteration $l$ is erased. Clearly,
\begin{align}\label{eq:deini}
  x^{(1)}_j =\epsilon, \quad j=1,2.
\end{align}
In the same way let $y^{(l)}_j$ be the probability that a message from a check node to a variable node 
on an edge of type $j$ in iteration $l$ is erased. This probability is
\begin{align}\label{eq:decheck}
  y_j^{(l)} & = 1 - \rho^{(j)}(1-x_j^{(l)}), \quad j=1,2.
\end{align}
Using this we can write down the following recursions for $x^{(l)}_j$:
\begin{align}
  \label{eq:2d DE1} x_1^{(l+1)} & = \epsilon \lambda^{(1)}(y_1^{(l)},y_2^{(l)})\\
  \label{eq:2d DE2} x_2^{(l+1)} & = \epsilon \lambda^{(2)}(y_1^{(l)},y_2^{(l)}).
\end{align}

We denote the binary entropy function by 
\begin{equation*}
h(x) \triangleq -x \log_2(x) -(1-x) \log_2(1-x).   
\end{equation*}

The indicator variable $\openone_{\{S\}}$ corresponding to a statement $S$ is given by 
\[
\openone_{\{S\}} = \left\{ \begin{array}{cc}
1 & \textnormal{if }S\textnormal{ is True,}\\
0 & \textnormal{Otherwise.}
\end{array}
\right.
\]
By $\coef{\sum_i F_i D^i}{D^j}$ we mean the coefficient of $D^j$ in the formal power sum $\sum_i F_i D^i$, i.e. $\coef{\sum_i F_i D^i}{D^j} = F_j$.

In the next section, we show how the degree distribution of a two edge
type LDPC ensemble can be chosen such that it has the same density
evolution recursion as that of a given standard LDPC ensemble. We also
numerically optimize the degree distribution of two edge type LDPC 
ensembles and show that we can approach points on the boundary of the
achievable rate-equivocation region.

\section{Design and Optimization}\label{sec:desandopt}
As the density evolution recursion is a two dimensional recursion for two edge type 
LDPC ensembles, it is difficult to analyze. Thus we look for degree distributions which reduce the
two dimensional recursion to a single dimension. This will enable us to use density evolution 
recursion for standard LDPC ensembles over the BEC, which has been very well studied. 
In the following theorem, we accomplish this 
task. 

\begin{thm}\label{thm:mainthm}
Let $(\lambda, \rho)$ be a standard LDPC degree distribution with design rate
$R$ and threshold $\epsilon^{\star}$ over the BEC. Then the following
assignment,
\begin{align}
  \rho^{(1)}(x) = \rho^{(2)}(x) &= \rho(x), \label{eq:rhoass} \\
  \lambda_{ll}^{(1)} =  \lambda_{ll}^{(2)}& = \lambda_{2l}, \label{eq:ass1}\\
  \lambda_{l l+1}^{(1)} = \lambda_{l+1 l}^{(2)}& = \frac{l}{2l+1} \lambda_{2l + 1}, \label{eq:ass2}\\
  \lambda_{l+1 l}^{(1)} = \lambda_{l l+1}^{(2)}& = \frac{l+1}{2l+1} \lambda_{2l + 1}, \label{eq:ass3} \\
  \lambda_{l_1 l_2}^{(1)} = \lambda_{l_1 l_2}^{(2)} & = 0, \quad |{l_1 - l_2}| > 1, \label{eq:ass4}
\end{align}
ensures that the two edge type LDPC ensemble $\left\{\lambda^{(1)},\lambda^{(2)}, \rho^{(1)}, \rho^{(2)}\right\}$ also has design rate $R$ and 
threshold $\epsilon^\star$.
\end{thm}

\begin{IEEEproof}
Assume that we choose $\lambda^{(1)},\lambda^{(2)}, \rho^{(1)}$, and $\rho^{(2)}$ such that 
(\ref{eq:rhoass})  and the following relation 
\begin{align}
  \label{eq:l constr.} \lambda^{(1)}(x,x) = \lambda^{(2)}(x,x) = \lambda(x).
\end{align}
is satisfied. Note that since
\begin{align*}
  \lambda^{(j)}(x,x) & = \sum_{l_1, l_2} \lambda^{(j)}_{l_1 l_2} x^{l_1 + l_2 - 1} \\
  & = \sum_{k} \left(\sum_{l_1 + l_2 = k} \lambda^{(j)}_{l_1 l_2}\right) x^{k-1} ,
\end{align*}
(\ref{eq:l constr.}) implies 
\begin{align}\label{eq:sumlambda12}
  \sum_{l_1 + l_2 = k} \lambda^{(1)}_{l_1 l_2} = \sum_{l_1 + l_2 = k} \lambda^{(2)}_{l_1 l_2} \quad \forall k. 
\end{align}
From the density evolution recursion for two edge type LDPC ensembles given in
(\ref{eq:deini}), (\ref{eq:decheck}), (\ref{eq:2d DE1}), and (\ref{eq:2d DE2}),
we see that (\ref{eq:rhoass}) ensures that $y_1^{(l)} = y_2^{(l)}$ whenever
$x_1^{(l)} = x_2^{(l)}$ and (\ref{eq:l constr.}) ensures $x_1^{(l+1)} =
x_2^{(l+1)}$ whenever $y_1^{(l)} = y_2^{(l)}$. Since $x_j^{(1)} = \epsilon$, by induction we see 
that $x_1^{(l)}=x_2^{(l)}$ and $y_1^{(l)}=y_2^{(l)}$ for $l \geq 1$. Thus we can reduce  
the two dimensional density evolution recursion to the one dimensional density evolution 
recursion for standard LDPC ensemble  
\begin{align}
  \label{eq:1d DE}
  x^{(l+1)} = \epsilon \lambda(1 - \rho(1-x^{(l)})),
\end{align}
where $\lambda(x) = \sum_k \lambda_k x^{k-1}$, 
\begin{align}\label{eq:sumlambda}
  \lambda_k = \sum_{l_1+l_2 = k} \lambda^{(1)}_{l_1 l_2}, 
\end{align}
and we have dropped the subscript of $x$ as $x_1^{(l)}=x_2^{(l)}$. 
Note that by (\ref{eq:ass1}), (\ref{eq:ass2}), (\ref{eq:ass3}), and 
(\ref{eq:ass4})
\begin{equation}
  \frac{\lambda^{(1)}_{l_1 l_2}}{l_1} = \frac{\lambda^{(2)}_{l_1 l_2}}{l_2} \quad \forall l_1,l_2. \label{eq:match4}
\end{equation}
This ensures that (\ref{eq:match3}) is fulfilled.

We now show that (\ref{eq:ass1}), (\ref{eq:ass2}), (\ref{eq:ass3}), and 
(\ref{eq:ass4}) guarantees that $\lambda^{(1)}(x,x) = \lambda^{(2)}(x,x) = \lambda(x)$. Then
the two dimensional density evolution recursion becomes the one dimensional
recursion in (\ref{eq:1d DE}) and the two type edge ensemble will have the
same threshold as the standard LDPC ensemble.  We have
\begin{align*}
  \lambda^{(1)}(x,x)  = &\sum_{l_1, l_2}\lambda^{(1)}_{l_1 l_2} x^{l_1 + l_2 -1}, \\
   \stackrel{\textrm{(a)}}{=} &\sum_{l} \brc{\lambda^{(1)}_{l l+1} x^{2l} + \lambda^{(1)}_{ll} x^{2l-1} + \lambda^{(1)}_{l+1 l} x^{2l}}, \\
   \stackrel{\textrm{(b)}}{=} &\sum_{l} \brc{\frac{l}{2l+1} \lambda_{2l + 1}  x^{2l} + \lambda_{2l}x^{2l-1}} + \\
  & \sum_{l} \frac{l+1}{2l+1} \lambda_{2l + 1} x^{2l},\\
   = &\sum_{l} \brc{\lambda_{2l + 1}  x^{2l} + \lambda_{2l}x^{2l-1}}, \\
   = &\lambda(x),
\end{align*}
where (a) is due to (\ref{eq:ass4}) and (b) is due to (\ref{eq:ass1})--(\ref{eq:ass3}).
The proof for $\lambda^{(2)}(x,x)$ is done in the same way.

We now show that the design rate of the resulting two edge type LDPC ensemble is the same 
as the design rate of the given standard LDPC ensemble. 
The design rate of the two edge type ensemble is 
\begin{align*}
  R_{\textrm{des}} = 1 - (m_1 + m_2)/n  
\end{align*}
where $m_j$ is the number of parity checks of type $j$ and $n$ is the number of
variable nodes. If we let $d_{\textrm{avg}}$ denote the average check node
degree (same for both the types because of (\ref{eq:rhoass})) and count the number 
of type $j$ edges in two different ways, we get
\begin{equation*}
  n\sum_{l_1,l_2} l_j \Lambda_{l_1l_2} = m_j d_{\textrm{avg}}, \quad j=1, 2. 
\end{equation*}
or
\begin{align*}
  \frac{m_j}{n} & = \frac{\sum_{l_1, l_2} l_j \Lambda_{l_1 l_2}}{d_{\textrm{avg}}}, \\
  & \stackrel{\textrm{(a)}}{=} \frac{1}{d_{\textrm{avg}}} \frac{\sum_{l_1, l_2} l_j \frac{\lambda^{(j)}_{l_1 l_2}}{l_j}}{ \sum_{l_1, l_2} \frac{\lambda^{(j)}_{l_1 l_2}}{l_j}}, \\
  & \stackrel{\textrm{(b)}}{=} \frac{1}{d_{\textrm{avg}}}\frac{1}{\sum_{l_1, l_2} \frac{\lambda^{(j)}_{l_1 l_2}}{l_j}},
\end{align*}
where (a) is due to (\ref{eq:match3}) and (b) follows since the $\lambda^{(1)}_{l_1 l_2}$ sum to 1.
The design rate then becomes
\begin{align*}
  R_{\textrm{des}} & = 1 - (m_1 + m_2)/n, \\
  & = 1 - \frac{1}{d_{\textrm{avg}}} \left(\frac{1}{\sum_{l_1, l_2} \frac{\lambda^{(1)}_{l_1 l_2}}{l_1}} + \frac{1}{\sum_{l_1, l_2} \frac{\lambda^{(2)}_{l_1 l_2}}{l_2}}\right), \\
  & \stackrel{\textrm{(a)}}{=} 1 - \frac{2}{d_{\textrm{avg}}} \left(\frac{1}{\sum_{l_1, l_2} \frac{\lambda^{(1)}_{l_1 l_2}}{l_1}} \right),\\
  & \stackrel{\textrm{(b)}}{=} 1 - \frac{2}{d_{\textrm{avg}}}\left(\frac{1}{\sum_{l} \brc{\frac{\lambda_{2l+1}}{2l+1}  + \frac{\lambda_{2l}}{l} + \frac{\lambda_{2l+1}}{2l+1}}}\right), \\
  & = 1 - \frac{1}{d_{\textrm{avg}}} \frac{1}{\sum_l \brc{\frac{\lambda_{2l+1}}{2l+1} + \frac{\lambda_{2l}}{2l}}}, \\
  & = 1 - \frac{1}{d_{\textrm{avg}}} \frac{1}{\sum_l \frac{\lambda_{l}}{l}},
\end{align*}
where (a) is due to (\ref{eq:match4}) and (b) follows using (\ref{eq:ass1}) -
(\ref{eq:ass4}). Since this expression is the same as the design rate of the
standard LDPC ensemble $(\lambda,\rho)$, we have shown that the two edge type
LDPC ensemble has design rate $R$. This completes the proof of the theorem. 
\end{IEEEproof}
To compute the threshold achievable on the main channel, we need to compute the threshold
of the ensemble of parity-check matrices $H_1$ corresponding to type one edges. The ensemble of matrices  
$H_1$ is a standard LDPC ensemble and its degree distribution can be easily
calculated from the degree distribution of the two edge type ensemble. Hence we
can easily compute its threshold. 

Since all capacity approaching sequences of degree distributions have some
degree two variable nodes, because of (\ref{eq:ass1}) we see that our
construction will have some degree one variable nodes in the matrix $H_1$. 
This means that the threshold over the main channel will
be zero. To get around this problem we use linear programming methods to find
good degree distributions for two edge type LDPC ensembles based on their 
two dimensional density evolution recursion.  

First we optimize the degree distribution of $H_1$ for the main channel using
the methods described in \cite{RiU08} and obtain a good ensemble
$(\Lambda^{(1)},\Gamma^{(1)})$. 

For a given two edge type ensemble we can find the corresponding one edge type
ensemble for $H_1$ by summing over the second index, since the
fraction of variable nodes with $l_1$ outgoing type one edges is given by
$\sum_{l_2} \Lambda_{l_1 l_2}$. To fix the degree distribution of $H_1$ we then
impose the constraint
\begin{align*}
  \sum_{l_2} \Lambda_{l_1 l_2} = \Lambda^{(1)}_{l_1} \textrm{ for all } l_1.
\end{align*}

For successful decoding we further impose the two constraints $x_1^{(l+1)} \leq
x_1^{(l)}$ and $x_2^{(l+1)} \leq x_2^{(l)}$ which can be written as
\begin{align*}
  x_1 &\geq \epsilon \lambda^{(1)}(y_1,y_2) \\
  & = \epsilon \sum_{l_1,l_2} \lambda^{(1)}_{l_1 l_2} y_1^{l_1 - 1} y_2^{l_2} \\
  & = \epsilon \sum_{l_1,l_2} \frac{l_1 \Lambda_{l_1,l_2}}{\sum_{k_1,k_2} k_1 \Lambda_{k_1,k_2}} y_1^{l_1 - 1} y_2^{l_2},
\end{align*}
where we have used (\ref{eq:match1}) in the last step, and $y_1,y_2$ are given by
\begin{align*}
  y_j = 1 - \rho_j(1-x_j), j=1,2.
\end{align*}
This simplifies to the linear constraint
\begin{align}
\label{eq:decon1}  0 \leq \sum_{l_1, l_2} l_1 (x_1 -  \epsilon y_1^{l_1 - 1} y_2^{l_2}) \Lambda_{l_1 l_2}.
\end{align}
The corresponding constraint for $x_2$ is
\begin{align}
 \label{eq:decon2} 0 \leq \sum_{l_1, l_2} l_1 (x_2 -  \epsilon y_1^{l_1} y_2^{l_2-1}) \Lambda_{l_1 l_2}.
\end{align}
The design rate can be written as
\begin{align*}
  R_{\textrm{des}} = 1 - \frac{\sum_{l_1, l_2} l_1 \Lambda_{l_1 l_2}}{\sum_{l_1} l_1 \Gamma^{(1)}_{l_1}} - 
\frac{\sum_{l_1, l_2} l_2 \Lambda_{l_1 l_2}}{\sum_{l_2} l_2 \Gamma^{(2)}_{l_2}},
\end{align*}
where the term $\frac{\sum_{l_1, l_2} l_1 \Lambda_{l_1 l_2}}{\sum_{l_1} l_1
\Gamma^{(1)}_{l_1}}$ is a constant because of the fixed degree distribution of $H_1$.
If $\Gamma^{(2)}$ is fixed we see that maximizing the design rate is the same
as minimizing $\sum_{l_1, l_2} l_2 \Lambda_{l_1 l_2}$. Thus we end up with the
following linear program, which we will solve iteratively:
\begin{align}\label{eq:linprog_obj}
  \textrm{minimize } \sum_{l_1, l_2} l_2 \Lambda_{l_1 l_2}
\end{align}
 subject to
 \begin{align}
  &\sum_{l_2} \Lambda_{l_1 l_2} = \Lambda^{(1)}_{l_1}, \ l_1 = 2, \ldots, I \label{eq:linprog_cons1}\\
  &\sum_{l_1, l_2} l_1 (x_1(k) -  \epsilon y_1(k)^{l_1 - 1} y_2(k)^{l_2}) \Lambda_{l_1 l_2}  \geq 0, \  k = 1,\ldots,K\label{eq:linprog_cons2}\\
  &\sum_{l_1, l_2} l_1 (x_2(k) -  \epsilon y_1(k)^{l_1} y_2(k)^{l_2-1}) \Lambda_{l_1 l_2} \geq 0, \ k = 1,\ldots,K,\label{eq:linprog_cons3}
\end{align}
where $I$ is the largest degree in $\Lambda^{(1)}(x)$. Since the constraints (\ref{eq:decon1}) and (\ref{eq:decon2}) correspond to infinitely many constraints we replace them by the first $K$ steps of the density evolution path followed by the degree distribution used in the previous iteration. Thus the points $\{x_1(k),
x_2(k)\}_{k=1}^K$ are chosen by generating a distribution $\Lambda_0$ and then
running the density evolution recursion
\begin{align}
  x_1^{(1)} &= x_2^{(1)} = \epsilon, \label{eq:linprog_de1}\\
  x_1^{(l+1)}& = \epsilon \lambda_0^{(1)}(y_1^{(l)},y_2^{(l)}), \label{eq:linprog_de2}\\
  x_2^{(l+1)} &= \epsilon \lambda_0^{(2)}(y_1^{(l)},y_2^{(l)}), \label{eq:linprog_de3}
\end{align}
$K$ times. The program is then solved repeatedly, each time updating $\{x_1(k),
x_2(k)\}_{k=1}^K$. This process is repeated several times for different check
node degree distributions $\Gamma^{(2)}$ until there is negligible improvement 
in rate. The complete optimization procedure is summarized in the following steps. 
\begin{enumerate}
\item Find an optimized degree distribution $(\Lambda^{(1)},\Gamma^{(1)})$ of 
$H_1$ for the main channel using the methods described in \cite{RiU08}. Fix 
a check node degree distribution $\Gamma^{(2)}$ corresponding to type two edges. 
\item Choose a two edge type variable node degree distribution $\Lambda$ which 
satisfies (\ref{eq:linprog_cons1}). 
\item Generate $K$ density evolution points $\{x_1(k), x_2(k)\}_{k=1}^K$ by using 
(\ref{eq:linprog_de1}), (\ref{eq:linprog_de2}), and  (\ref{eq:linprog_de3}). 
\item Solve the linear program given by (\ref{eq:linprog_obj}), (\ref{eq:linprog_cons1}), 
 (\ref{eq:linprog_cons2}), and (\ref{eq:linprog_cons3}). 
\item Repeat Step $3$) and Step $4$) until there is negligible improvement in rate.   
\end{enumerate}
As mentioned before, we repeat the optimization procedure for several $\Gamma^{(2)}$. 
A good choice of $\Gamma^{(2)}$ is either regular or with two different degrees.  

We now present some optimized degree distributions obtained by this
method. We use the following degree distribution
\begin{stdseq}
\label{stdseq:05}
\begin{align*}
  \Lambda^{(1)}(x) =& \ 0.5572098 x^2 + 0.1651436 x^3 + 0.07567923 x^4\\
&+0.0571348 x^5  + .043603 x^7 + 0.02679802 x^8 \\
&+0.013885518 x^{13} + 0.0294308 x^{14} + 0.02225301 x^{31}\\
&+0.00886105x^{100},\\
\Gamma^{(1)}(x) = & \ 0.25 x^9 + 0.75 x^{10}
\end{align*}
\end{stdseq}
\noindent as the ensemble $(\Lambda^{(1)},\Gamma^{(1)})$ for the main channel. It has
rate 0.498826, threshold 0.5, and multiplicative gap to capacity $(1-\epsilon -
R_{\textrm{des}})/(1-\epsilon) = 0.00232857$. We use it to obtain two optimized
degree distributions, one for $\epsilon_w =0.6$ and one for $\epsilon_w =
0.75$.

The degree distribution for the ensemble optimized for BEC-WT($0.5, 0.6$) is given by
\begin{tetseq}
\label{tetseq:05_06}
  \begin{align*}
    \Lambda(x,y) = &\  0.463846 x^2 + 0.0814943 x^2y + 0.0118691 x^2y^2 \\
    & +0.14239x^3 + 0.0201658 x^3 y + 0.00258812 x^3 y^2 \\
    &+0.0292241 x^4 + 0.0464551 x^4 y + 0.0564162 x^5 \\
    &+0.000718585 x^5 y + 0.0436039 x^7y\\
    &+ 0.0258926 x^8 y + 0.000905503 x^8y^2 \\
    &+ 0.00631474x^{13}y^2 + 0.00757076x^{13}y^5 \\
    &+ 0.011051x^{14}y + 0.0173718 x^{14}y^2 \\
    &+ 0.00100807x^{14}y^5 + 0.00240762x^{31} \\
    &+ 0.0012626x^{31}y^4 + 0.0185828 x^{31}y^5\\
    &+ 0.000326117 x^{100}y^4 + 0.00383319x^{100}y^{17}\\
    & + 0.00470174 x^{100}y^{18},\\
    \Gamma^{(1)}(x) & = 0.25x^9 + 0.75x^{10},\\
    \Gamma^{(2)}(x) & = x^6.
  \end{align*}
\end{tetseq}

This ensemble has design rate $0.39893$, threshold $0.6$, and the multiplicative gap to
capacity is $0.00267632$. The rate $R_{ab}$ from Alice to Bob is 0.099906 bits per channel use (b.p.c.u.)
and $R_e$, the equivocation of Eve, is $0.0989137$ b.p.c.u. 
However both $R_{ab}$ is very close to the secrecy 
capacity $C_S = 0.1$ b.p.c.u., and $R_{e}$ is very close to $R_{ab}$.  

The degree distribution for the ensemble optimized for BEC-WT($0.5, 0.75$)
is given by
\begin{tetseq}
\label{tetseq:05_075}
\begin{align*}
\Lambda(x,y) =& \ 0.367823 x^2 + 0.166244 x^2 y + 0.0231428 x^2 y^2 \\
&+ 0.125727x^3 + 0.0394166 x^3y+ 0.00286773 x^4\\
&+ 0.0728115x^4y + 0.0571348x^5y \\ 
&+0.0300989x^7y^2+ 0.013505x^7y^3 \\
&+    0.0196622x^8y^3 +    0.00713582x^8y^4\\
&+     0.000565918x^{13}y^2+     0.0133196x^{13}y^5\\
&+     0.0149732x^{14}y^2+     0.0132215x^{14}y^5 \\
&+     0.0012361x^{14}y^6+     0.00490831x^{31}y^8\\
&+     0.0173447x^{31}y^9+     0.00130606x^{100}y^{17}\\
&+     0.00498932x^{100}y^{30}+ 0.00256567x^{100}y^{31},\\
\Gamma^{(1)}(x) =& \ 0.25x^9 +0.75 x^{10}, \\
\Gamma^{(2)}(x) =& \ 0.25x^4 + 0.75 x^5.
\end{align*}
\end{tetseq}
This ensemble has design rate $0.248705$ and threshold $0.75$. The multiplicative gap to
capacity is $0.00518359$.  The rate $R_{ab}$ from Alice to Bob is $0.250131$ b.p.c.u.  
 and $R_e$, the equivocation of Eve, is $0.248837$ b.p.c.u. Note that the secrecy capacity 
$C_s$ for this channel is $0.25$ b.p.c.u. Thus the obtained point is slight to the right and 
below of point B in Figure \ref{fig:rate_equ}. 

As mentioned earlier, computing the equivocation of Eve is not as straightforward as 
computing the reliability on the main channel. In the next section we show how to
compute the equivocation of Eve by generalizing the methods from \cite{MMU08}
to two edge type LDPC codes.

\section{Preliminary Results for Computation of Equivocation}\label{sec:compequ}
In order to compute the average equivocation of Eve over the erasure pattern
and ensemble of codes, we generalize the MMU method of \cite{MMU08} to two edge
type LDPC codes. In \cite{MMU08}, the equivocation of standard LDPC ensembles
for point-to-point communication over BEC($\epsilon$) was computed. More
precisely, let $\ul{\tilde{X}}$ be a randomly chosen codeword of a randomly
chosen code $G$ from the standard LDPC ensemble. Let $\ul{\tilde{X}}$ be transmitted over 
BEC($\epsilon$) and let $\ul{\tilde{Z}}$ be the channel output. Then the MMU
method computes
\begin{equation}\label{eq:norm_ave_cond_entropy}
\lim_{n \to \infty} \frac{\mathbb{E}\brc{H_G(\ul{\tilde{X}}|\ul{\tilde{Z}})}}{n}, 
\end{equation}
where $H_G(\ul{\tilde{X}}|\ul{\tilde{Z}})$ is the conditional entropy of the transmitted 
codeword given the channel observation for the code $G$ and we do the averaging over 
the ensemble. Note
that we need not average over the codewords as the analysis can be carried out
under the assumption that the all-zero codeword is transmitted \cite[Chap.
3]{RiU08}. The MMU method is described below.  
\begin{enumerate}[\settowidth{\IEEElabelindent}{}] 
\item Consider decoding the all-zero codeword using the peeling decoder \cite[pp. 115]{RiU08}, which is 
described in the following. 
\begin{enumerate}
\item Initially, remove all the known (not erased from the channel) variable nodes
and the edges connected to them. Now remove all the degree zero check nodes. 
\item Pick a degree one check node. Declare its neighboring variable node to be 
known. Remove all the edges connected to this variable node. Remove all the 
 degree zero check nodes.  
\item If there are no degree one check nodes, then go to the next step. Otherwise, 
repeat the previous step.
\item Output the remaining graph which is called the {\it residual graph}. 
\end{enumerate}  
\item The peeling decoder gets stuck in the largest stopping set contained in
the set of erased variable nodes \cite{RiU08}. Thus the residual graph is the
subgraph induced by this stopping set. The residual graph is again a code whose
codewords are compatible with the erasure set.
\item The degree distribution of the residual graph and its edge connections
are random variables. It was shown in \cite{LMSS01} that if the erasure
probability is above the BP threshold, then almost surely the residual graph
has a degree distribution close to the {\it average residual degree
distribution}. The average residual degree distribution can be computed by the
asymptotic analysis of the  peeling decoder. Also, conditioned on the degree
distribution of the residual graph, the induced probability distribution is
uniform over all the graphs with the given degree distribution.  This implies
that almost surely a residual graph is an element of the standard LDPC ensemble 
with degree distribution equal to the average residual degree distribution, which 
we refer to as the {\it residual ensemble}.
\item The normalized expectation of the conditional entropy given in 
 (\ref{eq:norm_ave_cond_entropy}) can be determined from  
the average rate of the residual ensemble.  One can easily compute 
the design rate of the residual ensemble from its degree distribution.  However, the
design rate is only a lower bound on the average rate. A criterion was derived in 
\cite{MMU08}, which, when satisfied, guarantees that the average rate is equal
to the design rate. If the average rate is equal to the design rate, then the
 normalized expectation of the conditional entropy can be determined from the design rate of 
the residual ensemble. 
\end{enumerate}

For transmission over the BEC-WT($\epsilon_m, \epsilon_w$), to compute the equivocation 
of Eve $H(\ul{S}|\ul{Z})$, 
we write $H(\ul{S}, \ul{X}|\ul{Z})$ in two different ways using the chain rule and obtain 
\begin{multline}\label{eq:equiv_chainrule}
H(\ul{X}|\ul{Z}) +  H(\ul{S}|\ul{X},\ul{Z}) = H(\ul{S}|\ul{Z}) +  H(\ul{X}|\ul{S},\ul{Z}). 
\end{multline}
By noting that $H(\ul{S}|\ul{X},\ul{Z})=0$ and substituting it in (\ref{eq:equiv_chainrule}),
 we obtain 
\begin{equation}
\label{eq:equiv_eve}
   \frac{H(\ul{S}|\ul{Z})}{n} = \frac{H(\ul{X}|\ul{Z})}{n} - \frac{H(\ul{X}|\ul{S},\ul{Z})}{n}.
\end{equation}
In the following two subsections we show how the normalized averages of 
$H(\ul{X}|\ul{Z})$ and $H(\ul{X}|\ul{S},\ul{Z})$ can be computed. The next subsection deals 
with  $H(\ul{X}|\ul{Z})$. 

\subsection{Computing the Normalized $H(\ul{X}|\ul{Z})$}
In the following lemma we show that the average of $\lim_{n \to \infty} H(\ul{X}|\ul{Z})/n$ can be
computed by the MMU method. 
\begin{lem}\label{lem:hxzviaMMU}
Consider transmission over the BEC-WT($\epsilon_m, \epsilon_w$) using the syndrome
encoding method with a two edge type LDPC code $H = \begin{bmatrix} H_1 \\ H_2
\end{bmatrix}$, where the dimensions of $H$, $H_1$, and $H_2$ are $n (1-R) \times n$, 
$n (1-R_1) \times n$, and $n (R_1-R) \times n$ respectively. 
 Let $\ul{S}$ be a randomly chosen message from Alice for Bob and $\ul{X}$ be the 
transmitted vector which is a randomly chosen solution of $H \ul{X} =
\begin{bmatrix} \ul{0} \\ \ul{S}\end{bmatrix}$. Let $\ul{Z}$ be the channel
observation of the wiretapper Eve. Consider a point-to-point communication
set-up over BEC($\epsilon_w$) using a standard LDPC code $H_1$. Let $\ul{\hat{X}}$
be a randomly chosen transmitted codeword of the code given by $H_1$, i.e. $\ul{\hat{X}}$ is a randomly 
chosen solution of $H_1 \ul{X}=\ul{0}$. Further let $\ul{\hat{Z}}$ be the channel
output. Then
\[
	H\brc{\ul{X}|\ul{Z}} = H\brc{\ul{\hat{X}}|\ul{\hat{Z}}}.
\]
\end{lem}
\begin{IEEEproof}
We prove the lemma by showing that $(\ul{X}, \ul{Z})$ and $(\ul{\hat{X}}, \ul{\hat{Z}})$
have the same joint distribution. Clearly, $P(\ul{Z}=\ul{z}|\ul{X}=\ul{x}) =
P(\ul{\hat{Z}}=\ul{z}|\ul{\hat{X}}=\ul{x})$ as transmission takes place over
BEC($\epsilon_w$) in both the cases. Now 
\begin{align} 
P(\ul{X}=\ul{x}) & = \sum_{\ul{s}} P\brc{\ul{X}=\ul{x}, \ul{S}=\ul{s}}, \nonumber\\
	& \stackrel{(a)}{=} \frac{1}{2^{n (R_1-R)}} \sum_{\ul{s}} P\brc{\ul{X}=\ul{x}|\ul{S}=\ul{s}}, \nonumber\\
	& = \frac{1}{2^{n (R_1-R)}} \sum_{\ul{s}} \frac{1}{2^{n R}} \openone_{\{H_1 \ul{x} = \ul{0}\}} \openone_{\{H_2 \ul{x} = \ul{s}\}}, \nonumber\\
	& \stackrel{(b)}{=} \frac{\openone_{\{H_1 \ul{x} = \ul{0}\}}}{2^{n R_1}},\label{eq:priorX} 
\end{align}
where (a) follows from the uniform a priori distribution on $\ul{S}$ and (b) follows because for a
fixed $\ul{x}$,  
\[
\sum_{\ul{s}} \openone_{\{H_2 \ul{x} = \ul{s}\}} = 1.
\]  
Now the a priori distribution of $\ul{\hat{X}}$ is also the RHS of (\ref{eq:priorX}). This is because 
$\ul{\hat{X}}$ is a randomly chosen solution of $H_1 \ul{\hat{X}}=\ul{0}$. This proves the lemma. 
\end{IEEEproof}
From Lemma \ref{lem:hxzviaMMU}, we see that when we consider transmission over the
BEC-WT($\epsilon_m, \epsilon_w$) using the two edge type LDPC ensemble
$\{\Lambda, \Gamma^{(1)}, \Gamma^{(2)}\}$, we can compute the average of
$\lim_{n \to \infty} H(\ul{X}|\ul{Z})/n$ by applying the MMU method to
the standard LDPC ensemble $\{\Lambda^{(1)}, \Gamma^{(1)}\}$ for transmission
over the BEC($\epsilon_w$). We formally state this in the following theorem.  
\begin{thm}\label{thm:hxzviaMMU}
Consider transmission over the BEC-WT($\epsilon_m, \epsilon_w$) using a randomly chosen code $G$ 
from the two edge type 
LDPC ensemble $\{\Lambda, \Gamma^{(1)}, \Gamma^{(2)}\}$ and the coset encoding method. 
Let $\ul{X}$ be the transmitted word and $\ul{Z}$ be the wiretapper's observation. 

Consider a point-to-point communication setup for transmission over
BEC($\epsilon_w$) using a randomly chosen code $\hat{G}$ from the standard LDPC
ensemble $\{\Lambda^{(1)}, \Gamma^{(1)}\}$.  Let $\hat{\ul{X}}$ be a randomly chosen
transmitted codeword and $\hat{\ul{Z}}$ be the channel output.  Let $\{\Omega,
\Phi\}$ (from the node perspective), be the average residual degree distribution\footnote{$\Omega$ corresponding to the variable node degree distribution, and $\Phi$ corresponding to the check node degree distribution.}
of the residual ensemble given by the peeling decoder and let
$R^r_{\textnormal{des}}$ be the design rate of the average residual ensemble
$\{\Omega, \Phi\}$. If almost every element of the average residual ensemble
$\{\Omega, \Phi\}$ has its rate equal to the design rate
$R^r_{\textnormal{des}}$,  then
\begin{multline}
\lim_{n \to \infty} \frac{\mathbb{E}\brc{H_G(\ul{X}|\ul{Z})}}{n} = \lim_{n \to \infty} \frac{\mathbb{E}\brc{H_{\hat{G}}(\ul{\hat{X}}|\ul{\hat{Z}})}}{n} \\ 
   =  \epsilon_w \Lambda^{(1)}\brc{1-\rho^{(1)}(1-x)} R^r_\textnormal{des}, \label{eq:hxzviaMMU} 
\end{multline}
where $x$ is the fixed point of the density evolution recursion for $\{\Lambda^{(1)}, \Gamma^{(1)}\}$ initialized 
with erasure probability $\epsilon_w$, and $\rho^{(1)}$ is the check node degree distribution of $H_1$ from the edge perspective.
\end{thm}
{\it Remark:} Note that the condition that almost every element of the average residual
ensemble $\{\Omega, \Phi\}$ has its rate equal to the design rate can be
verified by using  \cite[Lem. 3.22]{RiU08} or \cite[Lem. 7]{MMU08}.
\begin{IEEEproof}
The first equality in (\ref{eq:hxzviaMMU}) is the result of Lemma  \ref{lem:hxzviaMMU}. 
The second equality of (\ref{eq:hxzviaMMU}) follows from  \cite[Thm. 10]{MMU08}. 
The factor $\epsilon_w \Lambda^{(1)}\brc{1-\rho^{(1)}(1-x)}$, which is the ratio of the block length 
of the average residual ensemble $\{\Omega, \Phi\}$ to the initial ensemble $\{\Lambda^{(1)}, \Gamma^{(1)}\}$,  
takes care of the fact that we are normalizing $H_G(\ul{X}|\ul{Z})$ by the block-length of the 
initial ensemble  $\{\Lambda^{(1)}, \Gamma^{(1)}\}$. 
\end{IEEEproof}

In the following section we generalize the MMU method to two edge type LDPC ensembles in order to 
compute $H(\ul{X}|\ul{S},\ul{Z})$. 

\subsection{Computing Normalized $H(\ul{X}|\ul{S},\ul{Z})$ by Generalizing the MMU method to 
the Two Edge Type LDPC Ensembles}
Similarly to Lemma \ref{lem:hxzviaMMU}, in the following lemma we show that
computing $H(\ul{X}|\ul{S},\ul{Z})$ for BEC-WT($\epsilon_m, \epsilon_w$) using 
the coset encoding method and two edge type LDPC ensemble $\{\Lambda, \Gamma^{(1)},
\Gamma^{(2)}\}$ is equivalent to computing the equivocation of the same  ensemble 
for point-to-point communication over BEC($\epsilon_w$). 
\begin{lem}\label{lem:hxszviaMMU}
Consider transmission over the BEC-WT($\epsilon_m, \epsilon_w$) using the syndrome
encoding method with a two edge type LDPC code $H = \begin{bmatrix} H_1 \\ H_2
\end{bmatrix}$, where the dimensions of $H$, $H_1$, and $H_2$ are $n (1-R) \times n$, 
$n (1-R_1) \times n$, and $n (R_1-R) \times n$ respectively. 
 Let $\ul{S}$ be a randomly chosen message from Alice for Bob and $\ul{X}$ be the 
transmitted vector which is a randomly chosen solution of $H \ul{X} =
\begin{bmatrix} \ul{0} \\ \ul{S}\end{bmatrix}$. Let $\ul{Z}$ be the channel
observation of the wiretapper Eve. 

Consider a point-to-point communication set-up for transmission over the BEC($\epsilon_w$) 
using a two edge type LDPC code $H = \begin{bmatrix} H_1 \\ H_2\end{bmatrix}$. Let 
$\ul{\hat{X}}$ be the transmitted codeword which is a randomly chosen solution of 
$H \ul{X}=\ul{0}$ and $\ul{\hat{Z}}$ be the channel output. Then 
\[
H(\ul{X}|\ul{S},\ul{Z}) \stackrel{\textnormal{(a)}}{=} H(\ul{X}|\ul{S}=\ul{0}, \ul{Z})  \stackrel{\textnormal{(b)}}{=} H(\ul{\hat{X}}|\ul{\hat{Z}}). 
\]
\end{lem}
\begin{IEEEproof}
Equality (b) is obvious. To prove equality (a), note that for a solution $\ul{x}$ of 
$H \ul{x} = \begin{bmatrix} \ul{0} \\ \ul{s}\end{bmatrix}$ we can write 
$\ul{x}=\ul{x}'\oplus\ul{x}_{\ul{s}}$, where $H \ul{x}'=\ul{0}$ and 
$H \ul{x}_{\ul{s}} = \begin{bmatrix} \ul{0} \\ \ul{s}\end{bmatrix}$. Let $\ul{z}$ be a specific received vector and let $\ul{z}'$ be the vector that has the same erased positions as $\ul{z}$ and is equal to the corresponding position in $\ul{x}'$ in the unerased positions. The proof is completed by noting that 
\begin{multline}
P(\ul{X}=\ul{x}, \ul{Z}=\ul{z}|\ul{S}=\ul{s}) = \\
P(\ul{X}=\ul{x}', \ul{Z}=\ul{z}'|\ul{S}=\ul{0}). 
\end{multline}
\end{IEEEproof}
Thus from Lemma \ref{lem:hxszviaMMU} we see that $H(\ul{X}|\ul{S},\ul{Z})$ can be computed by generalizing the MMU method to two edge type LDPC ensembles. In the next section we 
accomplish this task.

\section{MMU Method for Two Edge Type LDPC Ensembles}\label{sec:mmu_tet}

The peeling decoder described in Step 1 of the MMU method and its termination  
described in Step 2 is the same for two edge type LDPC ensembles. The proof of
Step 3 of the MMU method for two edge type LDPC ensembles is the same as that for 
standard LDPC ensembles. We state it in the following two lemmas.
\begin{lem}\label{lem:resgraphuniform}
Consider transmission over the BEC($\epsilon_w$) using the two edge type LDPC ensemble 
$\{\Lambda, \Gamma^{(1)}, \Gamma^{(2)}\}$ and decoding using the peeling decoder. 
Let $G$ be a random residual graph. Conditioned on the event that $G$ has 
degree distribution $\{\Omega, \Phi^{(1)}, \Phi^{(2)}\}$, it is equally likely 
to be any element of the two edge type ensemble $\{\Omega, \Phi^{(1)}, \Phi^{(2)}\}$. 
\end{lem}
\begin{IEEEproof}
The proof is the same as for the standard LDPC ensemble \cite{LMSS01imp}. However, for completeness the proof is given in Appendix \ref{sec:resgraphuniform}.
\end{IEEEproof}
 
\begin{lem}\label{lem:resdegconc}
Consider transmission over the BEC($\epsilon_w$) using the two edge type LDPC ensemble 
$\{\Lambda, \Gamma^{(1)}, \Gamma^{(2)}\}$ and decoding using the peeling decoder. 
Let $\{\Omega, \Phi^{(1)}, \Phi^{(2)}\}$ be the average residual degree distribution. Let 
$\{\Omega_G, \Phi^{(1)}_G, \Phi^{(2)}_G\}$ be the residual degree distribution of a random 
residual graph $G$. Then, for any $\delta > 0$
\[
\lim_{n \to \infty} P\left\{ d\brc{\brc{\Omega, \Phi^{(1)}, \Phi^{(2)}}, \brc{\Omega_G, \Phi^{(1)}_G, \Phi^{(2)}_G}} \geq \delta\right\} = 0.
\]
The distance $d(\cdot, \cdot)$ is the $L_1$ distance 
\begin{multline*}
d\brc{\brc{\Omega, \Phi^{(1)}, \Phi^{(2)}}, \brc{\tilde{\Omega}, \tilde{\Phi}^{(1)}, \tilde{\Phi}^{(2)}}} = \\
\sum_{i_1 i_2} |\Omega_{i_1 i_2} - \tilde{\Omega}_{i_1 i_2}| +  \sum_{j_1} |\Phi^{(1)}_{j_1} - \tilde{\Phi}^{(1)}_{j_1}| 
+ \sum_{j_2} |\Phi^{(2)}_{j_2} - \tilde{\Phi}^{(2)}_{j_2}|.
\end{multline*}
\end{lem}
\begin{IEEEproof}
The proof is very similar to the proof for the standard LDPC ensemble given in \cite[Thm. 3.106]{RiU08}. We provide an outline of the 
proof in Appendix \ref{sec:resdegconc}.
\end{IEEEproof}

In the following lemma we compute the average residual degree distribution of two edge type LDPC ensembles. 
\begin{lem}
\label{lem:res_dist}
Consider transmission over the BEC($\epsilon_w$) using the two edge type LDPC ensemble 
$\{\Lambda, \Gamma^{(1)}, \Gamma^{(2)}\}$ and decoding using the peeling decoder. 
  Let $(x_1,x_2)$ be the fixed points of $(\ref{eq:2d DE1})$ and 
$(\ref{eq:2d DE2})$ when initialized with channel erasure probability $\epsilon_w$. 
Let $y_j = 1 - \rho^{(j)}(1-x_j), \ j = 1,2$, where $\rho^{(j)}$ is the degree distribution 
of check nodes of type $j$ from the edge perspective. Then the average residual degree 
distribution $\{\Omega, \Phi^{(1)}, \Phi^{(2)}\}$ is  given by
  \begin{align*}
    \Omega(z_1,z_2) =&\  \epsilon_w \Lambda(z_1 y_1,z_2 y_2),\\
    \Phi^{(j)}(z) =&\  \Gamma^{(j)}(1-x_j + x_j z) - x_j z \Gamma'^{(j)}(1-x_j)\\
    &- \Gamma^{(j)}(1-x_j), \quad j=1,2,
  \end{align*}
where $\Gamma'^{(j)}(x)$ is the derivative of $\Gamma^{(j)}(x)$. Note that the degree distributions are normalized with respect to the number of variable 
(check) nodes in the original graph. 
\end{lem}
\begin{IEEEproof}
The proof follows by the analysis of the peeling decoder for general multi-edge type LDPC ensembles 
in \cite{HiW10}. However, as we are only interested in two edge type LDPC 
ensembles, the proof also follows from the analysis for the standard LDPC case \cite{LMSS01imp}. 
\end{IEEEproof}

Lemma \ref{lem:resgraphuniform}, \ref{lem:resdegconc}, and \ref{lem:res_dist}
generalize Step 3 of the MMU method for two edge type LDPC ensembles. The
key technical task in extending Step 4 to two edge type LDPC ensembles is to
derive a criterion, which when satisfied, guarantees that almost every code in
the residual ensemble has its rate equal to the design rate. The rate is equal
to the normalized logarithm of the total number of codewords. However, as the
average of the logarithm of the total number of codewords is hard to compute, we
compute the normalized logarithm of the average of the total number of codewords. By
Jensen's inequality this is an upper bound on the average rate. More precisely, let 
$N$ be the total number of codewords corresponding to a randomly chosen code. Then, 
by Jensen's inequality
\[
\lim_{n \to \infty} \frac{\mathbb{E}(\log_2(N))}{n} \leq \lim_{n \to \infty} \frac{\log_2\brc{\mathbb{E}(N)}}{n}
\]

If this upper bound
is equal to the design rate, then by the same arguments as in \cite[Lem. 7]{MMU08} 
we can show that almost every code in the ensemble has its rate equal to the design rate. 
In the following lemma we derive the average of the total number of codewords of a 
two edge type LDPC ensemble. 
 
\begin{lem}\label{lem:num_codewords}
Let $N$ be the total number of codewords of a randomly chosen code from 
 the two edge type LDPC ensemble $(\Lambda,\Gamma^{(1)},\Gamma^{(2)})$. Then the 
average of $N$ over the ensemble is given by
\begin{multline*}
    \mathbb E (N) =  \\ \sum_{E_1=0,E_2=0}^{n \Lambda_1'(1,1), n \Lambda_2'(1,1)} \coef{\prod_{l_1,l_2}(1+u_1^{l_1}u_2^{l_2})^{n\Lambda_{l_1,l_2}}}{u_1^{E_1}u_2^{E_2}} \times \\
\frac{\coef{\prod_{r_1,r_2}q_{r_1}(v_1)^{\frac{n\Lambda_1'(1,1)}{\Gamma'^{(1)}(1)}\Gamma^{(1)}_{r_1}}q_{r_2}(v_2)^{\frac{n\Lambda_2'(1,1)}{\Gamma'^{(2)}(1)}\Gamma^{(2)}_{r_2}}}{v_1^{E_1}v_2^{E_2}}}{{n\Lambda_1'(1,1)\choose E_1}{n\Lambda_2'(1,1)\choose E_2}}, 
  \end{multline*}
where $\Lambda_j'(1,1) = \sum_{l_1, l_2} l_j \Lambda_{l_1, l_2}$, $\Gamma'^{(j)}(1)=\sum_{r_j} r_j \Gamma^{(j)}_{r_j}$,  $j \in \{1, 2\}$. 
The polynomial $q_r(v)$ is defined as 
\begin{align}
  q_r(v) = \frac{(1+v)^r+(1-v)^r}{2}.
\end{align}
\end{lem}
\begin{IEEEproof}
Let $\mathcal{W}(E_1, E_2)$ be the set of assignments of ones and zeros to the
variable nodes which result in $E_1$ (resp. $E_2$) type one (resp. type two) edges
connected to variable nodes assigned value one. Denote the cardinality of
$\mathcal{W}(E_1, E_2)$ by $|\mathcal{W}(E_1, E_2)|$. For an assignment $w$,
let $\openone_w$ be a random indicator variable which evaluates to one if $w$ is a
codeword of a randomly chosen code and zero otherwise. Let $N(E_1, E_2)$ be the number of 
codewords belonging to the set $\mathcal{W}(E_1, E_2)$. 
Then we have the following relationships
\begin{align}
N(E_1, E_2) & = \sum_{w \in \mathcal{W}(E_1, E_2)} \openone_w, \label{eq:ne1e2_indsum}\\
N & = \sum_{E_1=0,E_2=0}^{n \Lambda_1'(1,1), n \Lambda_2'(1,1)} N(E_1, E_2). \label{eq:n_ne1e2sum} 
\end{align}
(\ref{eq:ne1e2_indsum}) follows simply by checking if every word in the set
$\mathcal{W}(E_1, E_2)$ is a codeword. We obtain (\ref{eq:n_ne1e2sum}) by 
partitioning the set of codewords based on the number of type one and type two edges   
connected to variables assigned value one. By linearity of expectation we
obtain
\begin{align}
\mathbb{E}(N(E_1, E_2)) & = \sum_{w \in \mathcal{W}(E_1, E_2)} \mathbb{E}(\openone_w), \\
\mathbb{E}(N) & = \sum_{E_1=0,E_2=0}^{n \Lambda_1'(1,1), n \Lambda_2'(1,1)} \mathbb{E}(N(E_1, E_2)) \label{eq:averageN}.
\end{align}
From the symmetry of code generation, we observe that $\mathbb{E}(\openone_w)$, for 
$w \in \mathcal{W}(E_1, E_2)$ is independent of $w$. Thus we can fix $w$ to any one element of 
$\mathcal{W}(E_1, E_2)$ and obtain 
\begin{equation}\label{eq:NE1E2}
\mathbb{E}(N(E_1, E_2)) = |\mathcal{W}(E_1, E_2)| \textnormal{Pr}\brc{w \textnormal{ is a codeword}}. 
\end{equation}
Note that $|\mathcal{W}(E_1, E_2)|$ is given by 
\begin{align}\label{eq:cardW}
|\mathcal{W}(E_1,E_2)| = \coef{\prod_{l_1,l_2}(1+u_1^{l_1}u_2^{l_2})^{n \Lambda_{l_1,l_2}}}{u_1^{E_1}u_2^{E_2}}. 
\end{align}
To understand (\ref{eq:cardW}), note that when a variable node with type one degree $l_1$ and type two degree $l_2$ is assigned a one, 
it gives rise to $l_1$ (resp. $l_2$) type one (resp. type two) edges connected to a variable node 
assigned value one, and when it is assigned a zero it gives rise to no such edges. Thus the generating function of such a variable node to count the number of edges 
it gives rise to, which are connected to a variable node assigned one, is given by $1+u_1^{l_1} u_2^{l_2}$. 
Hence the overall generating function is given by $\prod_{l_1,l_2}(1+u_1^{l_1}u_2^{l_2})^{n \Lambda_{l_1,l_2}}$. 

We now evaluate the probability that an assignment $w$, $w \in \mathcal{W}(E_1, E_2)$, is a codeword, which is 
given by  
\begin{multline}\label{eq:probw}
\textnormal{Pr}\brc{w \textnormal{ is a codeword}} = \\
 \frac{\textnormal{Total number of graphs for which }w\textnormal{ is a codeword}}
{\textnormal{Total number of graphs}}.  
\end{multline}
Similar to the arguments for the standard LDPC ensemble in \cite{MMU08}, 
the total number of graphs for which $w$ is a codeword is given by 
\begin{multline}\label{eq:num_w_codeword}
E_1! E_2! (n\Lambda_1'(1,1)-E_1)! (n\Lambda_2'(1,1)-E_2)! \\
\coef{\prod_{r_1,r_2}q_r(v_1)^{\frac{n\Lambda_1'(1,1)}{\Gamma'^{(1)}(1)}\Gamma^{(1)}_{r_1}}q_r(v_2)^{\frac{n\Lambda_2'(1,1)}{\Gamma'^{(2)}(1)}\Gamma^{(2)}_{r_2}}}{v_1^{E_1}v_2^{E_2}}. 
\end{multline}
The factorial term $E_1!$ in (\ref{eq:num_w_codeword}) corresponds to the fact that given a graph for which $w$ is a codeword, we can permute 
the check node side position of the $E_1$ type one edges connected to a variable node assigned value one, 
and $w$ will be a codeword for the resulting graph. Similarly, we obtain the other factorial terms in 
(\ref{eq:num_w_codeword}). The generating function in (\ref{eq:num_w_codeword}) is the 
generating function to count the number of ways edges can be assigned on the check node side such that 
$w$ is a codeword \cite{RiU08}.

By noting that the total number of graphs is equal to $(n\Lambda_1'(1,1))!
(n\Lambda_2'(1,1))!$, and combining (\ref{eq:averageN}),  (\ref{eq:NE1E2}),
(\ref{eq:cardW}), (\ref{eq:probw}), and (\ref{eq:num_w_codeword})  we obtain
the expression for the average of the total number of codewords. 
\end{IEEEproof}
{\it Remark:} Note that in Lemma \ref{lem:num_codewords} we count the number of codewords 
which give rise to $E_1$ type one (resp. $E_2$ type two) edges which are connected to 
a variable node assigned value one. A related quantity is the weight distribution of a code which counts the number of 
codewords with a given weight. The average weight distribution of two edge type and 
more generally multi-edge type LDPC ensembles have been computed in \cite{IKSSS05, KADPS09}. \\
Let $(e_1,e_2) = (E_1/(n\Lambda'_1(1,1)),E_2/(n\Lambda'_2(1,1)))$, i.e. $e_j$ is $E_j$ normalized by the total number of type $j$ edges, $j=1,2$. In the following lemma we find the set of $(e_1,e_2)$ for which $\lim_{n \to \infty} |\mathcal{W}(e_1n\Lambda'_1(1,1),e_2n\Lambda'_2(1,1))| \neq 0.$
\begin{lem}
  \label{lem:e_region}
  Let $\mathcal{E}(n)$ be the set of $(e_1,e_2)$ such that
  \begin{equation}
    \label{eq:E_sigma}
    \coef{\prod_{l_1,l_2}(1+u_1^{l_1}u_2^{l_2})^{n \Lambda_{l_1,l_2}}}{u_1^{e_1n\Lambda'_1(1,1)} u_2^{e_2n\Lambda'_2(1,1)}}\neq 0.
  \end{equation}
Let $\mathcal{E} \triangleq \lim_{n \to \infty} \mathcal{E}(n)$. Then $\mathcal{E}$ is given by 
\begin{multline*}
\mathcal{E} = 
\left\{(e_1,e_2) : \right. \\ \left. \brc{\frac{\sum_{l_1,l_2} l_1 \Lambda_{l_1,l_2}
\sigma(l_1,l_2)}{\Lambda'_1(1,1)},\frac{\sum_{l_1,l_2} l_2 \Lambda_{l_1,l_2}
\sigma(l_1,l_2)}{\Lambda'_2(1,1)}}\right\}, 
\end{multline*}
where $0 \leq \sigma(l_1,l_2)  \leq 1$.
\end{lem}
\begin{IEEEproof}
The proof is given in Appendix \ref{sec:e_region}.
\end{IEEEproof}
In the next lemma we show that $\mathcal{E}$ as defined in Lemma \ref{lem:e_region} is the set enclosed between two piecewise 
linear curves. 
\begin{lem}\label{lem:e_region_2}
Let $\mathcal{E}$ be as defined in Lemma \ref{lem:e_region}. Then $\mathcal{E}$ is the subset of $[0,1]^2$ enclosed between two piecewise linear curves. Order the pairs $(l_1,l_2)$ for which $\Lambda_{l_1,l_2} > 0$ in decreasing order of $l_1/l_2$ and assume that there are $D$ distinct such values. Let 
\begin{equation}
  \label{eq:sigma_d}
  \sigma_d(l_1,l_2) =
  \begin{cases}
    1 & \textnormal{if $l_1/l_2$ takes the $d^{\textnormal{th}}$ largest possible value,} \\
    0 & \textnormal{otherwise,}
  \end{cases}
\end{equation}
and let
\begin{equation}
  \label{eq:e_d}
  p_d = (\frac{\sum_{l_1,l_2} l_1 \Lambda_{l_1,l_2}
\sigma_d(l_1,l_2)}{\Lambda'_1(1,1)},\frac{\sum_{l_1,l_2} l_2 \Lambda_{l_1,l_2}
\sigma_d(l_1,l_2)}{\Lambda'_2(1,1)}).
\end{equation}
Then $\mathcal E$ is the set above the piecewise linear curve connecting the points $\{(0,0), p_1, p_1 + p_2,\ldots,(1,1)\}$ and below the piecewise linear curve connecting the points $\{(0,0), p_D, p_D + p_{D-1},\ldots,(1,1)\}$, where addition of points $p_1+p_2$ is the point obtained by 
component wise addition of $p_1$ and $p_2$. 
\end{lem}
\begin{IEEEproof}
The proof is given in Appendix \ref{sec:e_region_2}.
\end{IEEEproof}

In the following theorem and its corollary, we present a criterion for two edge type LDPC
ensembles, which, when satisfied, guarantees that the actual rate is equal to the 
design rate. In order to state the theorem, we define the function $\theta(e_1, e_2)$, 
which is used to calculate  the difference between the growth rate of the average of the total number of 
codewords and the design rate.   
\begin{multline}\label{eq:deftheta}
  \theta(e_1,e_2) = \sum_{l_1,l_2} \Lambda_{l_1,l_2} \log_2 (1 + u_1^{l_2} u_2^{l_2}) - \Lambda_1'(1,1) e_1 \log_2 u_1\\
  - \Lambda_2'(1,1)e_2 \log_2 u_2 + \frac{\Lambda_1'(1,1)}{\Gamma^{'(1)}(1)} \sum_{r_1} \Gamma^{(1)}_{r_1} \log_2 q_{r_1}(v_1) \\
- \Lambda_1'(1,1)e_1\log_2 v_1 + \frac{\Lambda_2'(1,1)}{\Gamma^{'(2)}(1)} \sum_{r_2} \Gamma^{(2)}_{r_2} \log_2 q_{r_2}(v_2) \\
- \Lambda_2'(1,1)e_2\log_2 v_2 - \Lambda_1'(1,1)h(e_1) - \Lambda_2'(1,1)h(e_2)- R_{\textnormal{des}},
\end{multline}
where $u_1,u_2, v_1,$ and $v_2$ are positive solutions to the following equations
\begin{align}\label{eq:der1}
  \frac{v_1}{\Gamma^{(1)'}(1)} \sum_{r_1} r_1 \Gamma_{r_1}^{(1)}  \frac{(1+v_1)^{r_1-1}-(1-v_1)^{r_1-1}}{(1+v_1)^{r_1}+(1-v_1)^{r_1}} & = e_1, \\
\label{eq:der2} \frac{v_2}{\Gamma^{(2)'}(1)} \sum_{r_2} r_2 \Gamma_{r_2}^{(2)}  \frac{(1+v_2)^{r_2-1}-(1-v_2)^{r_2-1}}{(1+v_2)^{r_2}+(1-v_2)^{r_2}} & = e_2, \\
\frac{1}{\Lambda_1'(1,1)} \sum_{l_1,l_2} \Lambda_{l_1,l_2} l_1 \frac{u_1^{l_1}u_2^{l_2}}{1+u_1^{l_1}u_2^{l_2}} = e_1, \\ \label{eq:der4}
\frac{1}{\Lambda_2'(1,1)} \sum_{l_1,l_2} \Lambda_{l_1,l_2} l_2 \frac{u_1^{l_1}u_2^{l_2}}{1+u_1^{l_1}u_2^{l_2}} = e_2.
\end{align}
\begin{thm}
\label{thm:crit_irregular}
Consider the two edge type LDPC ensemble $(\Lambda,\Gamma^{(1)},\Gamma^{(2)})$ with design rate 
$R_{\textnormal{des}}$. Let $N$ be the total number of codewords of a randomly chosen code $G$ from 
this ensemble and let $R_G$ be the actual rate of the code $G$. Then
  \begin{align*}
    \lim_{n \to \infty} \frac{ \log_2(\mathbb E [N])}{n} = \sup_{(e_1,e_2) \in \mathcal{E}} \theta(e_1,e_2) + R_{\textnormal{des}},
  \end{align*}
where $\theta(e_1, e_2)$ is defined in (\ref{eq:deftheta}) and the set $\mathcal{E}$ is defined in Lemma \ref{lem:e_region}. 
\end{thm}
\begin{IEEEproof}
By (\ref{eq:averageN}), we have  
\begin{multline*}
\lim_{n \to \infty} \frac{ \log_2(\mathbb E [N])}{n} = \\ \sup_{(e_1, e_2) \in \mathcal{E}}
 \lim_{n \to \infty} \frac{\log_2(\mathbb E [N(e_1n \Lambda'_1(1,1),e_2 n
\Lambda'_2(1,1))])}{n}.
\end{multline*}
Using Stirling's approximation for the binomial coefficients and \cite[Theorem 2]{BuMi04} for
the coefficient growths in Lemma \ref{lem:num_codewords} we know that
\begin{multline}
  \lim_{n \to \infty} \frac{\log_2(\mathbb E [N(e_1n \Lambda'_1(1,1),e_2 n \Lambda'_2(1,1))])}{n} = \\
  \sup_{(e_1,e_2) \in \mathcal E} \inf_{u_1,u_2,v_1,v_2 > 0} \psi(e_1,e_2,u_1,u_2,v_1,v_2)
\end{multline}
where $\psi(e_1,e_2,u_1,u_2,v_1,v_2)$ is given by
\begin{multline}
\sum_{l_1,l_2} \Lambda_{l_1,l_2} \log_2 (1 + u_1^{l_2} u_2^{l_2}) - \Lambda_1'(1,1) e_1 \log_2 u_1\\
  - \Lambda_2'(1,1)e_2 \log_2 u_2 + \frac{\Lambda_1'(1,1)}{\Gamma^{'(1)}(1)} \sum_{r_1} \Gamma^{(1)}_{r_1} \log_2 q_{r_1}(v_1) \\
- \Lambda_1'(1,1)e_1\log_2 v_1 + \frac{\Lambda_2'(1,1)}{\Gamma^{'(2)}(1)} \sum_{r_2} \Gamma^{(2)}_{r_2} \log_2 q_{r_2}(v_2) \\
- \Lambda_2'(1,1)e_2\log_2 v_2 - \Lambda_1'(1,1)h(e_1) - \Lambda_2'(1,1)h(e_2).
\end{multline}
Further, the infimum of $\psi$ with respect to $u_1, u_2, v_1,$ and $v_2$  is given by solving the following 
saddle point equations
\begin{align}
  \frac{\partial \psi}{\partial u_1} =   \frac{\partial \psi}{\partial u_2} =   \frac{\partial \psi}{\partial v_1} =   
\frac{\partial \psi}{\partial v_2} = 0, 
\end{align}
which are equivalent to (\ref{eq:der1}) - (\ref{eq:der4}). 
\end{IEEEproof}
We now state the condition, which, when satisfied, guarantees that the actual rate is equal to the design rate. 
\begin{cor}
Let $\theta(e_1, e_2)$ be as defined in (\ref{eq:deftheta}). If $\sup_{(e_1,e_2) \in \mathcal{E}} \theta(e_1,e_2)=0$ i.e. if $\theta(1/2, 1/2) \geq \theta(e_1, e_2), \forall (e_1, e_2) \in 
\mathcal{E}$, then for any $\delta > 0$
\[
	\lim_{n \to \infty} P\brc{R_G \geq R_{\textnormal{des}} + \delta} = 0. 
\]
The set $\mathcal{E}$ is defined in Lemma \ref{lem:e_region}.
\end{cor}
\begin{IEEEproof}
From Theorem \ref{thm:crit_irregular}, $\mathbb{E}[N]=2^{n (R_{\textnormal{des}} + o(1))}$. Now from Markov's inequality, 
\begin{align*}
P\brc{R_G \geq R_{\textnormal{des}} + \delta} & = P\brc{N \geq 2^{n (R_{\textnormal{des}} + o(1) + \delta)}}, \\
                                              & \stackrel{(a)}{\leq} 2^{-n \delta},  
\end{align*}
where (a) follows from Markov's inequality. This proves the corollary. 
\end{IEEEproof}

Note that in general for a two edge type LDPC ensemble, in order to check if
the actual rate is equal to the design rate, we need to compute the maximum of
a two variable function over the set $\mathcal{E}$.  However, the set $\mathcal{E}$
is just a line for two edge type left regular LDPC ensembles. 
Thus we deal with 
the case of left regular LDPC ensembles in the following lemma. 
\begin{lem}\label{lem:crit_left_reg}
Consider the left regular two edge type LDPC ensemble $\{l_1, l_2, \Gamma^{(1)},
\Gamma^{(2)}\}$ with design rate $R_{\textnormal{des}}$. Let $N$ be the total
number of codewords of a randomly chosen code $G$ from this ensemble and $R_G$
be its actual rate. Then 
\begin{align*}
  \lim_{n \to \infty} \frac{ \log_2(\mathbb E [N])}{n} = \sup_{e \in (0,1)} \theta(e) + R_{\textnormal{des}}. 
\end{align*}
If $\sup_{e \in (0,1)} \theta(e) = 0$ i.e. if $\theta(1/2) \geq \theta(e), \forall e \in (0, 1)$, then 
 for any $\delta > 0$ 
\[
\lim_{n \to \infty} \mathbb{P}\brc{R_G > R_{\textnormal{des}} + \delta}= 0 
\]
The function $\theta(e)$ is defined as 
\begin{multline*}
 \theta(e) =
(1-l_1-l_2)h(e) + \frac{l_1}{\Gamma^{(1)'}(1)} \sum_r \Gamma^{(1)}_r \log q_r(v_1)\\
+ \frac{l_2}{\Gamma^{(2)'}(1)} \sum_r \Gamma^{(2)}_r \log q_r(v_2) - el_1 \log v_1 - e l_2 \log v_2-R_{\textnormal{des}}, 
\end{multline*}
where $v_1$ (resp. $v_2$) is the unique positive solution of (\ref{eq:der1}) (resp. (\ref{eq:der2})) with $e_1$ (resp. $e_2$) 
substituted by $e$ on the RHS. 
\end{lem}
\begin{IEEEproof}
Most of the arguments in this lemma are the same as those of Theorem \ref{thm:crit_irregular}, so we will omit them. 
First note that the cardinality of the set $\mathcal{W}(E_1,E_2)$, as defined in Lemma \ref{lem:num_codewords}, 
is given by  
\begin{align*}
  |\mathcal{W}(E_1,E_2)| &= \coef{(1+u_1^{l_1}u_2^{l_2})^n}{u_1^{E_1}u_2^{E_2}} \\
  & = \begin{cases} 0 & \frac{E_2}{l_2} \neq \frac{E_1}{l_1} \\
  {n \choose E_1/l_1} & \text{otherwise}\end{cases}
\end{align*}
Let $e = E_1/(nl_1) = E_2/(n l_2)$. By Stirling's approximation and the saddle point 
approximation for the coefficient terms \cite[pp. 517]{RiU08}, we obtain 
  \begin{align*}
    \lim_{n \to \infty} \frac{ \log_2(\mathbb E [N])}{n}  &=  \lim_{n \to \infty} \sup_{e \in (0,1)}\frac{\log_2(\mathbb E [N(enl_1,enl_2)])}{n}\\
    &= \sup_{e \in (0,1)} \inf_{v_1,v_2 > 0}  \psi(e,v_1,v_2),
  \end{align*}
where 
\begin{align*}
  \psi(e,v_1,v_2) =& (1-l_1-l_2) h(e) \\ 
 &+ \frac{l_1}{\Gamma^{(1)'}(1)} \sum_{r_1} \Gamma^{(1)}_{r_1} \log_2 q_{r_1}(v_1) - e \l_1 \log_2 v_1 \\
 &  +\frac{l_1}{\Gamma^{(2)'}(1)} \sum_r \Gamma^{(2)}_{r} \log_2 q_r(v_2) - e \l_2 \log_2 v_2.
\end{align*}
The saddle point equations are obtained by taking the partial derivatives of
$\psi$ with respect to $v_j, j \in \{1, 2\}$ and setting them equal to 0. These
equations are the same as (\ref{eq:der1}) (resp. (\ref{eq:der2})) with $e_1$ (resp.
$e_2$) substituted by $e$ on the RHS.
\end{IEEEproof}

{\it Remark:} Note that as in \cite{MMU08}, we can change the order of $\inf$
and $\sup$.  Taking the derivatives after changing the order gives a function
which is an upper bound on $\theta(e)$. The advantage of this upper bound is
that it can be computed without solving any saddle point equations. However as
opposed to the standard LDPC ensembles, for two edge type LDPC ensembles this upper
bound is not tight and does not provide a meaningful criterion to check if the
rate is equal to the design rate. 

The following two lemmas show that in the case of a left regular ensemble where
$\Gamma^{(1)}$ and $\Gamma^{(2)}$ both have only either odd or even degrees,
the function $\theta(e)$ attains its maximum inside the interval $[0, 1/2]$.
\begin{lem}\label{lem:checkodddeg}
Consider the left regular two edge type LDPC ensemble $\{l_1, l_2,
\Gamma^{(1)}, \Gamma^{(2)}\}$.  Let $\theta(e)$ be the function as defined in
Lemma \ref{lem:crit_left_reg}.  If both $\Gamma^{(1)}$ and $\Gamma^{(2)}$ are
such that both the type of check nodes only have odd degrees, then for $e >
1/2$  
\[
\theta(e) < \theta(1/2). 
\]
\end{lem}
\begin{IEEEproof}
The proof is given in Appendix \ref{sec:checkodddeg}. 
\end{IEEEproof}

\begin{lem}\label{lem:checkevendeg}
 Consider the left regular two edge type LDPC ensemble $\{l_1, l_2,
\Gamma^{(1)}, \Gamma^{(2)}\}$.  Let $\theta(e)$ be the function as defined in
Lemma \ref{lem:crit_left_reg}.  If both $\Gamma^{(1)}$ and $\Gamma^{(2)}$ are
such that both the type of check nodes only have even degrees, then for $e \in (0, 1/2)$  
\[
\theta(e) = \theta(1-e). 
\]
\end{lem}
\begin{IEEEproof}
The proof is given in Appendix \ref{sec:checkevendeg}.
\end{IEEEproof}

In the following theorem we state how we can compute the  
quantity $H(\ul{X}|\ul{S}, \ul{Z})$ appearing in (\ref{eq:equiv_eve}).

\begin{thm}\label{lem:cond_entr}
Consider transmission over the BEC-WT($\epsilon_m, \epsilon_w$) using a random code $G$ from 
the two edge type 
LDPC ensemble $\{\Lambda, \Gamma^{(1)}, \Gamma^{(2)}\}$ and the coset encoding method. 
Let $\ul{S}$ be the information word from Alice for Bob, $\ul{X}$ be the transmitted word,
 and $\ul{Z}$ be the wiretapper's observation. 

Also consider a point-to-point communication setup for transmission over the BEC($\epsilon_w$) 
using the two edge type LDPC ensemble $\{\Lambda, \Gamma^{(1)}, \Gamma^{(2)}\}$
Assume that the erasure probability $\epsilon_w$ is above the BP
threshold of the ensemble. Let $\{\Omega, \Phi^{(1)}, \Phi^{(2)}\}$ be the average residual ensemble 
resulting from the peeling decoder. Let $R^r_{\textnormal{des}}$ be the design rate of the 
residual ensemble $\{\Omega, \Phi^{(1)}, \Phi^{(2)}\}$. If $\{\Omega, \Phi^{(1)}, \Phi^{(2)}\}$ satisfies the 
condition of Theorem \ref{thm:crit_irregular}, i.e. if the design rate of the residual ensemble 
is equal to the rate then 
\begin{align}
 \label{eq:cond_ent} \lim_{n \to \infty}\frac{\mathbb{E}(H_G(\ul{X}|\ul{S}, \ul{Z}))}{n} & = \epsilon_w \Lambda(y_1, y_2) R^{r}_{\textnormal{des}}, 
\end{align}
where $x_1,x_2,y_1,$ and $y_2$ are the fixed points of the density evolution equations $(\ref{eq:2d DE1})$ and
$(\ref{eq:2d DE2})$ obtained when initializing them with $x_1^{(1)} = x_2^{(2)} = \epsilon_w$.
\end{thm}
\begin{IEEEproof}
From Lemma \ref{lem:hxszviaMMU}, we know that the conditional entropy in the point-to-point 
set-up is identical to $H(\ul{X}|\ul{S}, \ul{Z})$. The conditional entropy in the point-to-point 
case is equal to the RHS of (\ref{eq:cond_ent}). This follows from the same arguments as in 
\cite[Thm. 10]{MMU08}. The quantity $\epsilon_w \Lambda(y_1, y_2)$ on the RHS of (\ref{eq:cond_ent}) is the 
ratio of the number of variable nodes in the residual ensemble to that in the initial ensemble. 
\end{IEEEproof}

This gives us the following method to calculate the equivocation of Eve when using two edge type LDPC ensembles for the BEC-WT($\epsilon_m, \epsilon_w$) 
based on the coset encoding method.
\begin{enumerate}
\item If the threshold of the two edge type LDPC ensemble is lower than
$\epsilon_w$, calculate the residual degree distribution for the two edge type
LDPC ensemble for transmission over the BEC($\epsilon_w$).  Check that the rate
of this residual ensemble is equal to the design rate using Theorem
\ref{thm:crit_irregular}.  Calculate $H(\ul{X}|\ul{S},\ul{Z})$ using Theorem
\ref{lem:cond_entr}.  If the threshold is higher than $\epsilon_w$,
$H(\ul{X}|\ul{S},\ul{Z})$ is trivially zero. 
\item If the threshold of the standard LDPC ensemble induced by type one edges
is higher than $\epsilon_w$, calculate the residual degree distribution of this
ensemble for transmission over the BEC($\epsilon_w$).  Check that its rate is
equal to the design rate using \cite[Lemma 7]{MMU08}. Calculate
$H(\ul{X}|\ul{Z})$ using Theorem \ref{thm:hxzviaMMU}.  If the threshold is
higher than $\epsilon_w$, $H(\ul{X}|\ul{Z})$ is trivially zero. 
\item Finally calculate $H(\ul{S}|\ul{Z})$ using (\ref{eq:equiv_eve}).
\end{enumerate}
In the next section we demonstrate this procedure by computing the equivocation of Eve for various 
two edge type LDPC ensembles.

\section{Examples}\label{sec:examples}
\begin{example}
\label{ex:05_06_full_rate}
Consider using the ensemble defined by Standard LDPC degree distribution \ref{stdseq:05}, defined in Section \ref{sec:desandopt}, for transmission over the BEC-WT($0.5,0.6$) at rate $R_{ab}=0.498836$ b.p.c.u. (the full rate of the ensemble), without using the coset encoding scheme. Here every possible message $\ul{s}$ corresponds to a single codeword $\ul{x}$, and encoding and decoding is done as with a standard LDPC code. Since the threshold is $0.5$, Bob can decode with error probability approaching zero. The equivocation of Eve is given by $H(\ul{S}|\ul{Z}) = H(\ul{X}|\ul{Z})$ which can be calculated using the MMU method. In Figure \ref{fig:example1} we plot the function $\Psi_{\{\Omega^{(1)}, \Phi^{(1)}\}}(u)$ defined in 
\cite[Lemma 7]{MMU08} corresponding to the standard LDPC ensemble $\{\Omega^{(1)}, \Phi^{(1)}\}$, which is the average residual 
degree distribution of the ensemble induced by type one edges for transmission over the
 BEC($\epsilon_w$).
\begin{figure}
  \centering
  \includegraphics[width=0.75\columnwidth]{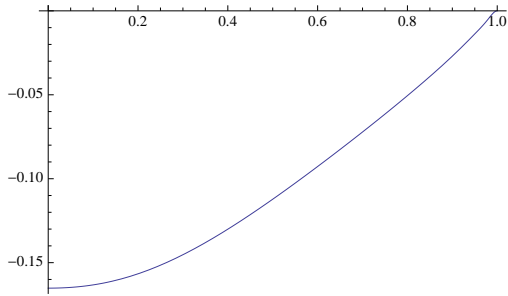}
  \caption{$\Psi_{\{\Omega^{(1)}, \Phi^{(1)}\}}(u)$ for Example \ref{ex:05_06_full_rate} and \ref{ex:05_06_Cs}.}
  \label{fig:example1}
\end{figure}
From \cite[Lemma 7]{MMU08}, if the maximum of $\Psi_{\{\Omega^{(1)}, \Phi^{(1)}\}}(u)$ over the unit
interval occurs at $u=1$, which holds in this case, the design rate of the residual graph is equal to the
actual rate. Thus we can calculate the average equivocation $\lim_{n \to \infty}
H(\ul{X}|\ul{Z})/n = 0.0989137$ b.p.c.u. Using this ensemble we can achieve the point
$(R_{ab},R_e) = (0.498836,0.0989137)$ in the rate-equivocation region
which is very close to the point C = $(0.5,0.1)$ in Figure \ref{fig:rate_equ}.
\end{example}

\begin{example}
\label{ex:05_06_Cs}
Now consider the two edge type ensemble defined by Two Edge Type Degree Distribution \ref{tetseq:05_06}, defined in Section \ref{sec:desandopt},
 for transmission over the BEC-WT($0.5,0.6)$ using the coset encoding
 scheme. Again Bob can decode since the threshold of the ensemble
 induced by type one edges is 0.5. Since the threshold of the two edge
 type ensemble is $0.6$ we get $H(\ul{X}|\ul{S},\ul{Z}) = 0$, and we
 get $H(\ul{S}|\ul{Z}) = H(\ul{X}|\ul{Z})$. The degree distribution of
 the type one edges is the same as the degree distribution in Example 1, so we again get 
$\lim_{n \to \infty} \mathbb{E}(H(\ul{X}|\ul{Z}))/n = 0.0989137$. Using this scheme we achieve the point  $(R_{ab},R_e) = (0.0999064,0.0989137)$ in the rate-equivocation region which is very close to point B = $(0.1,0.1)$ in Figure \ref{fig:rate_equ}.
\end{example}

\begin{example}
\label{ex:regular}
Consider transmission over the BEC-WT($0.429,0.75$) using the coset encoding scheme and the regular two edge type ensemble defined by
\begin{tetseq}
\begin{align}
  \Lambda(x,y) &= x^3y^3 \\
  \Gamma^{(1)}(x) & = x^6\\
\Gamma^{(2)}(x) & = x^{12}.
\end{align}
\end{tetseq}
The design rate of this ensemble is $0.25$ and the threshold is
$0.469746$. The threshold for the ensemble induced by type one edges
is $0.4294$, so it can be used for reliable communication if
$\epsilon_m < 0.4294$.

To calculate the equivocation of Eve, we first calculate
$\lim_{n \to \infty} H(\ul{X}|\ul{Z})/n$ by the MMU method. We calculate the average
residual degree distribution $\{\Omega^{(1)}, \Phi^{(1)}\}$ of the
ensemble induced by type one edges for erasure probability
$\epsilon_w$ and plot $\Psi_{\{\Omega^{(1)}, \Phi^{(1)}\}}(u)$ in
Figure \ref{fig:example3}. As in Examples 1 and 2, we see that it takes
its maximum at $u = 1$. Thus, by \cite[Lemma 7]{MMU08}, we obtain that
the conditional entropy is equal to the design rate of the residual
ensemble normalized with respect to the number of variable nodes in the original ensemble, i.e. $\lim_{n \to \infty} \mathbb{E}(H(\ul{X}|\ul{Z}))/n =
0.250124$ b.p.c.u.

We now calculate the average residual degree distribution
$(\Omega,\Phi^{(1)},\Phi^{(2)})$ of the two edge type ensemble
corresponding to erasure probability $\epsilon_w$ and plot the
function $\theta(e)$ defined in Lemma \ref{lem:crit_left_reg}. If
$\theta(e)$ is less than or equal to zero for $e \in [0,1]$, then the
rate of the residual ensemble is equal to the design rate by Lemma
\ref{lem:crit_left_reg}. Then we can calculate
$H(\ul{X}|\ul{S},\ul{Z})$ using Lemma \ref{lem:cond_entr}. In Figure
\ref{fig:example3} we see that $\sup_{e \in [0,1]} \theta(e) = 0$, and
we get $\lim_{n \to \infty} \mathbb{E}(H(\ul{X}|\ul{S},\ul{Z}))/n =
0.000124297$ b.p.c.u.

Finally, using (\ref{eq:equiv_eve}) we get $\mathbb{E}(H(\ul{S}|\ul{Z}))/n = 0.24999998$. We thus achieve the point $(R_{ab},R_e) = (0.25,0.24999998)$ in the rate-equivocation region. We see that we are very close to perfect secrecy. The reason that we are so far away from the secrecy capacity $C_s = 0.321$ is that the $(3,6)$ ensemble for the main channel is far from being capacity achieving.

\begin{figure}
  \centering
  \begin{minipage}{0.49\columnwidth}
    \includegraphics[width=\columnwidth]{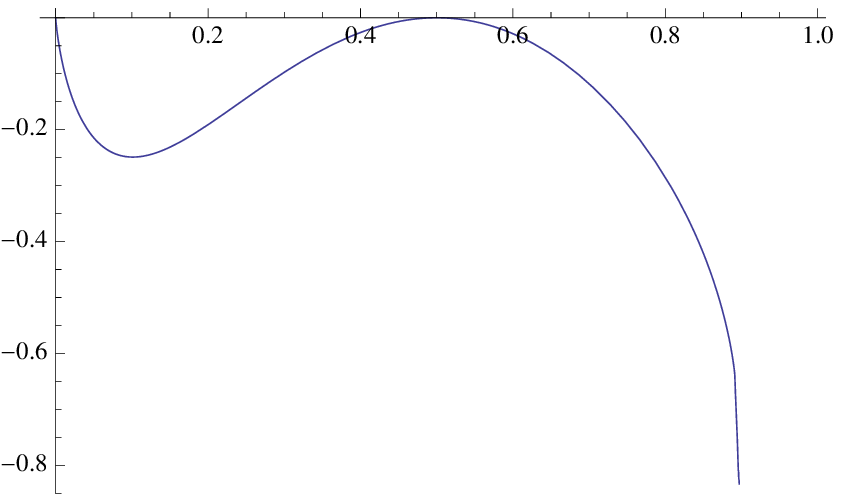}
  \end{minipage}
  \begin{minipage}{0.49\columnwidth}
    \includegraphics[width=\columnwidth]{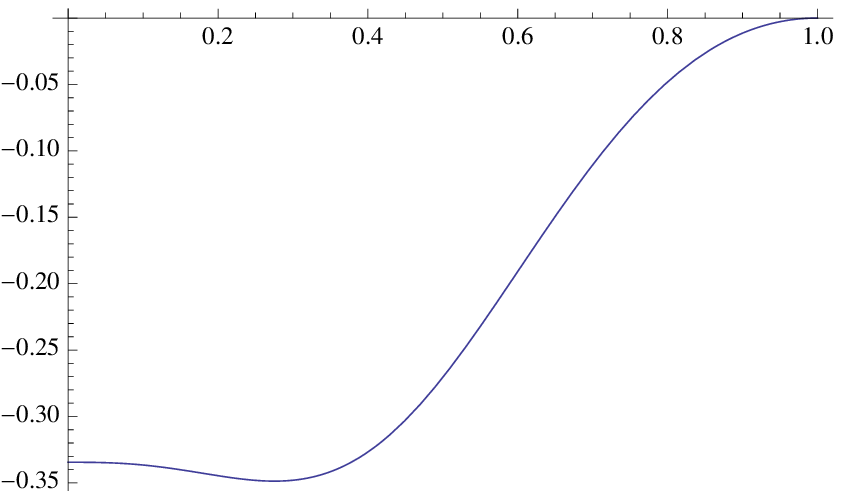}
  \end{minipage}
  \caption{$\theta(e)$ and $\Psi_{\{\Omega^{(1)}, \Phi^{(1)}\}}(u)$ for Example \ref{ex:regular}.}
  \label{fig:example3}
\end{figure}

\end{example}

\begin{example}
Consider the two edge type ensemble
\begin{tetseq}
\begin{align*}
    \Lambda(x,y) = &0.5572098 x^2y^3 + 0.1651436 x^3y^3 \\ 
    &+ 0.07567923 x^4y^3 +0.0571348 x^5y^3  \\ &+ .043603 x^7y^3 + 0.02679802 x^8y^3 \\
&+0.013885518 x^{13}y^3 + 0.0294308 x^{14}y^3 \\ &+ 0.02225301 x^{31}y^3
+0.00886105x^{100}y^3,
\end{align*}
\vspace*{-0.2in}
  \begin{align*}
\Gamma^{(1)}(x) = & \ 0.25 x^9 + 0.75 x^{10}, \\
\Gamma^{(2)}(x) = & \ x^{12}
  \end{align*}
\end{tetseq}
\noindent where the graph induced by type one edges has the same
degree distribution as Standard LDPC Degree Distribution
\ref{stdseq:05} and the graph induced by type two edges is $(3,12)$
regular. The rate of the overall ensemble is $0.248836$ and the rate from
Alice to Bob is $R_{ab} = 0.25$ b.p.c.u. Consider transmission over the
BEC-WT($0.5,0.751164$).

In Figure \ref{fig:example6}, we plot $\Psi_{\{\Omega^{(1)},
  \Phi^{(1)}\}}(u)$ for the residual ensemble $\{\Omega^{(1)},
\Phi^{(1)}\}$ induced by type one edges for transmission over
the BEC($\epsilon_w$). Since the maximum of $\Psi_{\{\Omega^{(1)},
  \Phi^{(1)}\}}(u)$ over the unit interval occurs at $u=1$ we obtain
by \cite[Lemma 7]{MMU08} that the rate is equal to the design rate for
this residual ensemble.  In Figure \ref{fig:example6} we plot
$\theta(e_1,e_2)$ for the residual ensemble
$(\Omega,\Phi^{(1)},\Phi^{(2)})$ of the two edge type LDPC ensemble
for transmission over the BEC($\epsilon_w$). Since the maximum of
$\theta(e_1,e_2)$ over the unit square is zero, we obtain by Theorem
\ref{thm:crit_irregular} that the rate is equal
to the design rate for this residual two edge type ensemble. In this
case we can calculate the equivocation of Eve and find it to be
$0.24999999$ b.p.c.u., which is very close to the rate. Thus this ensemble
achieves the point $(R,R_e) = (0.25,0.24999999)$ in the capacity-equivocation region in Figure \ref{fig:rate_equ}. Note that the secrecy capacity is $0.251164$ b.p.c.u.
\begin{figure}
  \centering
  \begin{minipage}{0.49\columnwidth}
    \includegraphics[width=\columnwidth]{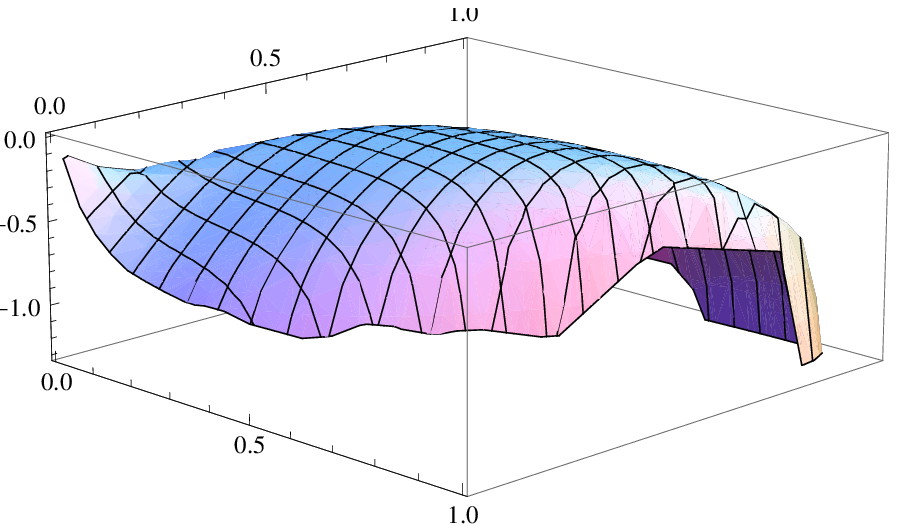}
  \end{minipage}
  \begin{minipage}{0.49\columnwidth}
    \includegraphics[width=\columnwidth]{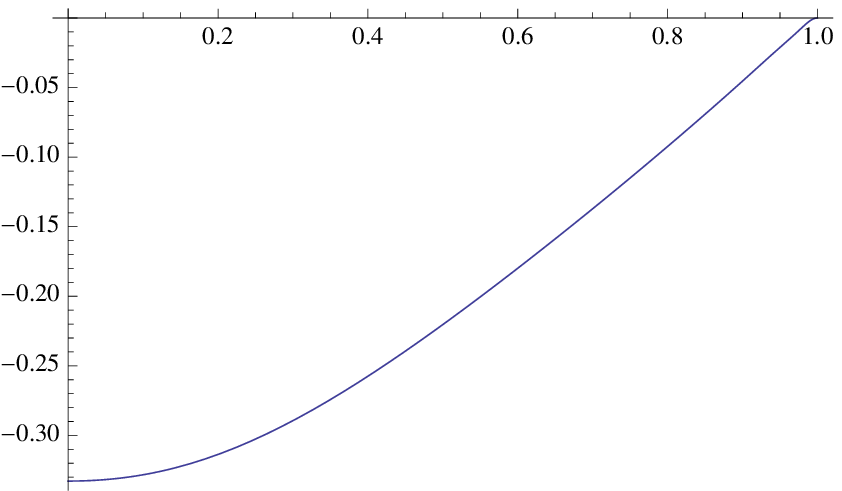}
  \end{minipage}
  \caption{$\theta(e_1,e_2)$ and $\Psi_{\{\Omega^{(1)}, \Phi^{(1)}\}}(u)$ for Example 4.}
  \label{fig:example6}
\end{figure}
\end{example}
These examples demonstrate that simple ensembles have very good  
 secrecy performance when the weak notion of secrecy is considered.  

%
%
%


\section{Conclusion}\label{sec:conc}

We consider the use of two edge type LDPC codes for the binary erasure wiretap
channel. The reliability performance can be easily measured using density
evolution recursion. We generalize the method of \cite{MMU08} to two edge type
LDPC codes in order to measure the security performance. We find that relative simple 
ensembles have very good secrecy performance. We have constructed a
capacity achieving sequence of two edge type LDPC ensembles for the BEC based on
capacity achieving sequences for the standard LDPC ensemble. However, this
construction introduces some degree one variable nodes in the ensemble for the main
channel, requiring an erasure free main channel.  We use linear programming
methods to find ensembles that operate close to secrecy capacity. However, as the 
underlying channel in our setup is a BEC, it is highly desirable to construct 
explicit sequences of secrecy capacity achieving ensembles. Due to the two dimensional 
recursion of density evolution for two edge type LDPC ensembles this is a much harder 
problem. In our opinion, this is one of the fundamental open problems in the setting 
of using sparse graph codes for transmission over the BEC-WT($\epsilon_m, \epsilon_w$).

\bibliographystyle{IEEEtran} 
\bibliography{IEEEabrv,secret}

\appendices 
\section{Proof of Lemma \ref{lem:resgraphuniform}}\label{sec:resgraphuniform}
\begin{proof}
Consider a residual graph $G$. Consider two type one edges $e_1$ and $e_2$ 
(the argument is the same for type two edges). Swap the check node side end points of 
$e_1$ and $e_2$. We denote the resulting graph by $G'$. The proof is completed by noting 
that the number of erasure patterns which result in $G$ are equal to the number of number 
of erasure patterns which result in $G'$. This is because if the variable nodes in $G$ form the largest 
stopping set in the erasure pattern then so do the variable nodes in $G'$.
\end{proof}
\section{Proof Outline of Lemma \ref{lem:resdegconc}}\label{sec:resdegconc}
\begin{proof}
The proof for the standard LDPC case uses the Wormald technique described in
\cite[App. C]{RiU08}.  Our proof is the same as that for the standard LDPC case except
that we have to keep track of the degree distribution of two different types of
edges. 

Assume that in the peeling decoder a degree one check node is chosen randomly
from the set of degree one check nodes. Let $G(t)$ be the residual graph after
the $t^{\tiny th}$ iteration of the peeling decoder. Let $V^{(1)}_{i_1 i_2}(t)$
(resp. $V^{(2)}_{i_1 i_2}(t)$) be the number of type one (resp. type 2) edges
which are connected to a variable node of degree $(i_1, i_2)$ in $G(t)$. 
For $j \in \{1, 2\}$, let $V^{(j)}(t)$ be the vector of number of type $j$ edges
of different degrees i.e.  $V^{(j)}(t)=\{V^{(j)}_{i_1 i_2}(t)\}_{i_1, i_2}$.
Let $C^{(1)}_i(t)$ (resp. $C^{(2)}_i(t)$) be the number of type one (resp. type
two) edges which are connected to type one (resp. type two) check nodes of
degree $i$ at time $t$. For $j \in \{1, 2\}$, let
$C^{(j)}(t)=\{C^{(j)}_i(t)\}_i$.  To show the concentration of the residual
degree distribution using the Wormald technique, we note that $\brc{V^{(1)}(t),
V^{(2)}(t), C^{(1)}(t), C^{(2)}(t)}$ is a Markov process.  The next requirement is
that the maximum possible change in $V^{(j)}_{i_1 i_2}(t)$ and $C^{(j)}_i(t)$ for
$j \in \{1, 2\}$, for all $(i_1, i_2)$ and for all $i$ after an iteration of the
peeling decoder should be bounded. This is true as all the degrees are finite.  
 The functions which describe the expected change in  
$V^{(j)}_{i_1 i_2}(t)$ and $C^{(j)}_i(t)$ are also Lipschitz continuous in $\brc{V^{(1)}(t)/n,
V^{(2)}(t)/n, C^{(1)}(t)/n, C^{(2)}(t)/n}$, where $n$ is the number of variable nodes. 
For example, as long as 
$C^{(1)}_1(t) + C^{(2)}_1(t) > 0$, for $j \in \{1, 2\}$
\begin{multline*}
\hspace{-0.1in} \mathbb{E}\left[ V^{(j)}_{i_1 i_2}(t+1)- V^{(j)}_{i_1 i_2}(t)|V^{(1)}(t), V^{(2)}(t), C^{(1)}(t), C^{(2)}(t)\right] 
\\ = - \frac{i_j V^{(j)}_{i_1 i_2}}{\sum_{l_1,l_2}  V^{(j)}_{l_1 l_2}}.  
\end{multline*}
The RHS of the previous equation is the same as that for the standard LDPC ensemble which has been shown to be
 Lipschitz continuous. 

The last required condition is that of initial concentration, i.e. the concentration condition 
should be satisfied at the beginning of the peeling decoder. This proof is the same as that for the 
standard LDPC ensemble given in \cite[App. C]{RiU08}. 
\end{proof}
\section{Proof of Lemma \ref{lem:e_region}}\label{sec:e_region}
\begin{proof}
The terms in the expansion of $\prod_{l_1,l_2}(1+u_1^{l_1}u_2^{l_2})^{n
\Lambda_{l_1,l_2}}$ have the form 
\[
u_1^{\sum_{l_1,l_2} l_1 k(l_1,l_2) \Lambda_{l_1,l_2}} u_2^{\sum_{l_1,l_2} l_2 k(l_1,l_2) \Lambda_{l_1,l_2}}, 
\]
where
$0 \leq k(l_1,l_2) \leq n$. If the coefficient of $u_1^{e_1n\Lambda'_1(1,1)}
u_2^{e_2n\Lambda'_2(1,1)}$ is non-zero, there exist $\{k(l_1, l_2)\}_{l_1, l_2}$ 
such that 
\begin{align*}
  \sum_{l_1,l_2} l_1 k(l_1,l_2) \Lambda_{l_1,l_2} = e_1 n \Lambda'_1(1,1)
\end{align*}
and
\begin{align*}
  \sum_{l_1,l_2} l_2 k(l_1,l_2) \Lambda_{l_1,l_2} = e_2 n \Lambda'_2(1,1)
\end{align*}
which is the same as
\begin{align*}
(e_1,e_2) = (\frac{\sum_{l_1,l_2} l_1 \Lambda_{l_1,l_2}
\sigma(l_1,l_2)}{\Lambda'_1(1,1)},\frac{\sum_{l_1,l_2} l_2 \Lambda_{l_1,l_2}
\sigma(l_1,l_2)}{\Lambda'_2(1,1)}), 
\end{align*}
where $0 \leq \sigma(l_1,l_2) = k(l_1,l_2)/n \leq 1$. When $n$ grows this is the same as (\ref{eq:E_sigma}).
\end{proof}
\section{Proof of Lemma \ref{lem:e_region_2}}\label{sec:e_region_2}
\begin{proof}
We show that $\mathcal{E}$ is the set between the two piecewise linear curves described in the statement of this lemma. We show this by varying the $\sigma(l_1,l_2)$ between $0$ and $1$ while trying to make the ratio $e_1/e_2$ as large as possible. Start by letting $\sigma(l_1,l_2) = 0$ if $l_1/l_2$ is not maximal, and letting $\sigma(l_1,l_2)$ increase to $1$ if $l_1/l_2$ is maximal. This traces out the line between $(0,0)$ and $p_1$, and clearly we can not have $(e_1,e_2)$ below this line for $(e_1,e_2) \in \mathcal E$. Then increase $\sigma(l_1,l_2)$ for $l_1,l_2$ such that $l_1/l_2$ takes the second largest value. This traces out the line between $p_1$ and $p_1 + p_2$ and again it is clear that we can not have $(e_1,e_2)$ below this line for $(e_1,e_2) \in \mathcal E$. We continue like this until we have $\sigma(l_1,l_2) = 1$ for all $l_1,l_2$, which corresponds to the point $(1,1)$.
The upper curve is obtained by reversing the order and starting with the line between $(0,0)$ and $p_D$.
\end{proof}
\section{Proof of Lemma \ref{lem:checkodddeg}}\label{sec:checkodddeg}
\begin{proof}
Take the derivative of $\theta(e)$ with respect to $e$ to get
\begin{align*}
  \frac{d \theta}{d e} = &(1-l_1 - l_2) \log\left(\frac{1-e}{e}\right) - l_1 \log v_1 - l_2 \log v_2 \\
  = &\log\left(\frac{1-e}{e}\right) - l_1 \log\left(\frac{(1-e)v_1}{e}\right)\\
  & - l_2 \log\left(\frac{(1-e)v_2}{e}\right).
\end{align*}
Using (\ref{eq:der1}) and (\ref{eq:der2}) we obtain
\begin{align*}
  \frac{1-e}{e} & = \frac{1-\frac{v_1}{\Gamma^{(1)'}(1)} \sum_{r_1} r_1 \Gamma_{r_1}^{(1)}  \frac{(1+v_1)^{r_1-1}-(1-v_1)^{r_1-1}}{(1+v_1)^{r_1}+(1-v_1)^{r_1}} }{\frac{v_1}{\Gamma^{(1)'}(1)} \sum_{r_1} r_1 \Gamma_{r_1}^{(1)}  \frac{(1+v_1)^{r_1-1}-(1-v_1)^{r_1-1}}{(1+v_1)^{r_1}+(1-v_1)^{r_1}}} \\
  & = \frac{\sum_{r_1} r_1 \Gamma_{r_1}^{(1)} \left(1 -  v_1 \frac{(1+v_1)^{r_1-1}-(1-v_1)^{r_1-1}}{(1+v_1)^{r_1}+(1-v_1)^{r_1}}\right)}{\sum_{r_1} r_1 \Gamma_{r_1}^{(1)}  v_1\frac{(1+v_1)^{r_1-1}-(1-v_1)^{r_1-1}}{(1+v_1)^{r_1}+(1-v_1)^{r_1}}} \\
  & = \frac{\sum_{r_1} r_1 \Gamma_{r_1}^{(1)} \frac{(1+v_1)^{r_1-1}+(1-v_1)^{r_1-1}}{(1+v_1)^{r_1}+(1-v_1)^{r_1}}}{\sum_{r_1} r_1 \Gamma_{r_1}^{(1)}  v_1\frac{(1+v_1)^{r_1-1}-(1-v_1)^{r_1-1}}{(1+v_1)^{r_1}+(1-v_1)^{r_1}}}
\end{align*}
or
\begin{align}\label{eq:r_odd}
  \frac{(1-e)v_1}{e} & =  \frac{\sum_{r_1} r_1 \Gamma_{r_1}^{(1)} \frac{(1+v_1)^{r_1-1}+(1-v_1)^{r_1-1}}{(1+v_1)^{r_1}+(1-v_1)^{r_1}}}{\sum_{r_1} r_1 \Gamma_{r_1}^{(1)}  \frac{(1+v_1)^{r_1-1}-(1-v_1)^{r_1-1}}{(1+v_1)^{r_1}+(1-v_1)^{r_1}}}.
\end{align}
We obtain a similar expression for $(1-e)v_2/e$.
Note that $v_j(e)$ are increasing functions of $e$ and $v_j(1/2) = 1$. Thus for $e>1/2, \ v_j > 1$ which together with (\ref{eq:r_odd}) implies $\frac{(1-e)v_j}{e} > 1$ when all $r$ are odd. This in turn implies that $\frac{d\theta}{de} < 0$ for $e>1/2$.
\end{proof}
\section{Proof of Lemma \ref{lem:checkevendeg}}\label{sec:checkevendeg}
\begin{proof}
  First we show that $v(1-e) = 1/v(e)$ if there are only even degrees. Let $v_j(e) = v$ and $1/v = \tilde v$. Then
  \begin{align*}
    e &= \frac{1/\tilde v}{\Gamma^{(j)'}(1)} \sum_{r} r \Gamma_{r}^{(j)}  \frac{(1+1/\tilde v)^{r-1}-(1-1/\tilde v)^{r-1}}{(1+1/\tilde v)^{r}+(1-1/\tilde v)^{r}} \\
    & = \frac{1}{\Gamma^{(j)'}(1)} \sum_{r} r \Gamma_{r}^{(j)}  \frac{(1+\tilde v)^{r-1}+(1-\tilde v)^{r-1}}{(1+\tilde v)^{r}+(1-\tilde v)^{r}}
  \end{align*}
and
\begin{align*}
  1-e & = 1 - \frac{v}{\Gamma^{(j)'}(1)} \sum_{r} r \Gamma_{r}^{(j)}  \frac{(1+v)^{r-1}-(1-v)^{r-1}}{(1+v)^{r}+(1-v)^{r}}\\
  & = \frac{1}{\Gamma^{(j)'}(1)}\sum_r \Gamma^{(j)}_r \left(1 - v \frac{(1+v)^{r-1}-(1-v)^{r-1}}{(1+v)^{r}+(1-v)^{r}}\right)\\
  & = \frac{1}{\Gamma^{(j)'}(1)} \sum_{r} r \Gamma_{r}^{(j)}  \frac{(1+v)^{r-1}+(1-v)^{r-1}}{(1+ v)^{r}+(1-v)^{r}}
\end{align*}
These two equations imply that $v(1-e) = 1/v(e)$. Now note that
\begin{align*}
  q_r(1/v) = \frac{q_r(v)}{v^r}
\end{align*}
for $r$ even, so 
\begin{align*}
   \theta(1-e) = &\ (1-l_1-l_2)h(1-e) + \frac{l_1}{\Gamma^{(1)'}(1)} \sum_r \Gamma^{(1)}_r \log \frac{q_r(v_1)}{v_1^r} \\
&+ \frac{l_2}{\Gamma^{(2)'}(1)} \sum_r \Gamma^{(2)}_r \log \frac{q_r(v_2)}{v_2^r} -(1-e) l_1 \log (1/v_1) \\ & - (1-e) l_2 \log (1/v_2)-R_{\textnormal{des}} \\
= &\ (1-l_1-l_2)h(1-e) + \frac{l_1}{\Gamma^{(1)'}(1)} \sum_r \Gamma^{(1)}_r \log q_r(v_1) \\
&- l_1 \log v_1+ \frac{l_2}{\Gamma^{(2)'}(1)} \sum_r \Gamma^{(2)}_r \log q_r(v_2) -l_2 \log v_2 \\ &+(1-e) l_1 \log (v_1) + (1-e) l_2 \log (v_2)-R_{\textnormal{des}} \\
= &\ \theta(e),
\end{align*}
using that 
\begin{equation*}
  \frac{l_j}{\Gamma^{(j)'}(1)} \sum_r \Gamma^{(j)}_r \log v_j^r = \frac{l_j}{\Gamma^{(j)'}(1)} \sum_r r \Gamma^{(j)}_r \log v_j = l_j \log v_j
\end{equation*}
in the second equality.
\end{proof}

\end{document}